\documentclass[review,12pt]{elsarticle}

\usepackage{lineno,hyperref}
\modulolinenumbers[5]
\usepackage[margin=3.0cm]{geometry}
\usepackage{amssymb}
\usepackage{amsthm}
\usepackage{amsmath}
\usepackage{amsfonts}
\usepackage[capitalise]{cleveref}
\usepackage{tcolorbox}
\usepackage{graphicx}
\usepackage{adjustbox}
\usepackage{caption}
\usepackage[labelformat=simple, position=b]{subcaption}

\usepackage{floatrow}
\usepackage{color}
\usepackage{booktabs}
\usepackage{multicol}
\usepackage{multirow}
\usepackage{tikz}
\biboptions{numbers,sort&compress}

\usepackage{xr}
\usepackage{soul,xcolor}
\setstcolor{red}
\setul{}{2pt}

\journal{Acta Materialia}

\begin{document}
\begin{frontmatter}

\title{Nucleation Kinetics in Al-Sm Metallic Glasses}

\author{L. Zhao}
\author{G.B. Bokas}
\author{J.H. Perepezko}
\author{I. Szlufarska\corref{I. Szlufarska}}

\cortext[I. Szlufarska]{Author to whom correspondence should be addressed}
\ead{szlufarska@wisc.edu}

\address{Department of Materials Science and Engineering, University of Wisconsin - Madison, United States}

\begin{abstract}
The isothermal nucleation kinetics in Al-Sm metallic glasses with low Sm concentrations ($x_\mathrm{Sm}$) was studied using molecular dynamics simulations in order to calculate time$-$temperature$-$transformation curves. The average delay time of Al nanocrystal nucleation was found to increase exponentially with $x_\mathrm{Sm}$, whereas the estimated critical cooling rate necessary to avoid crystallization decreases exponentially with $x_\mathrm{Sm}$. Sm solutes were found to suppress Al nucleation by increasing the attachment barrier and therefore by reducing the attachment frequency. The analysis shows that the attachment of Al to the evolving nucleus has the same characteristics as Al diffusion within the amorphous matrix and they both take place heterogeneously via collective movement of a group of Al atoms.
\end{abstract}

\begin{keyword}
Metallic glass \sep nucleation \sep delay time \sep attachment \sep molecular dynamics
\end{keyword}

\end{frontmatter}


\section{Introduction}
Metallic glasses (MGs) have attracted a growing interest since they were first reported in 1960 by Duwez \textit{et al.}~\cite{klement1960non}, due to their superior mechanical properties, better corrosion resistance and formability~\cite{greer2002wear, schuh2007mechanical, kumar2009nanomoulding, inoue2011recent}, as compared to their crystalline counterparts. In particular, Al-based MGs have a lower density and higher specific strength, and therefore have become promising candidates for applications as structural components~\cite{inoue1988glass, he1993unique}. Rapidly quenched Al-based MGs are usually characterized by a primary crystallization reaction upon heating which produces a high density of face-centered cubic (FCC) Al nanocrystals and consequently enhances the mechanical properties evidently~\cite{he1988synthesis, foley1996analysis, wilde1999glass}. For instance, it has been reported that the fracture strength of MGs with such nanocrystalline dispersions is 20 to 120\% higher than that of the pure amorphous phase alloys with the same compositions~\cite{chen1991mechanical, kim1991increase, kim1991ultrahigh}. 

The primary crystallization is of significant importance in understanding glass formation of alloys and there are two aspects of the primary crystallization that are still not well understood. One of them concerns unknown mechanisms underlying the effect of micro-alloying on the glass forming ability (GFA) of Al-based and other systems. This effect is known to be significant~\cite{huang2008primary, huang2008effects}. The second issue that needs to be elucidated is the nucleation kinetics in the primary crystallization~\cite{perepezko2000synthesis}, especially on the atomic level. To address these issues, here molecular dynamics (MD) simulations were carried out to investigate the isothermal nucleation reactions of Al-Sm binary MG. This glass has been reported in experiments to undergo the primary crystallization during annealing~\cite{wu2001glass, perepezko2000synthesis, perepezko2002amorphous, allen1998nanocrystal}. In this study, a particular attention is devoted to Sm effect on the nucleation kinetics and the atomic-level mechanism controlling the nucleation attachment. Simulation of the crystal growth after the nucleation is outside the scope of this study.

\section{Methods}
MD simulations are performed using the LAMMPS simulation package~\cite{plimpton1995fast}, based on a Finnis-Sinclair type semi-empirical potential developed for Al-Sm alloy by Mendelev~\textit{et al.}~\cite{mendelev2015development}. Although this potential has been fitted only to a limited number of properties during the development procedure, it has been shown to reproduce many properties that are relevant to the current study. These include the cohesive energy, melting temperature, and fusion enthalpy of pure Al, formation energies of Al-Sm crystal phases~\cite{mendelev2015development}, and icosahedral ordering during rapid solidification of Al-Sm alloys~\cite{bokas2016role}. In addition, this potential has been demonstrated to predict the same total and partial pair distribution functions as \textit{ab initio} MD simulations in Al$_{90}$Sm$_{10}$ liquid~\cite{mendelev2015development} and supercooled liquid~\cite{sun2016cooling}, and produce structure factors in a reasonable agreement with experimental measurements in Al$_{90}$Sm$_{10}$ MGs~\cite{sun2016cooling}. This potential is therefore generally suitable for simulating solidification/vitrification in Al-Sm system at low Sm concentrations.

In the simulations, a 9 $\times$ 9 $\times$ 9~nm$^3$ simulation box was used, that contained a total of 42,592 atoms, and the periodic boundary conditions were enforced in all three Cartesian directions. An isothermal-isobaric (NPT) ensemble is used in all the simulations and the temperature and pressure are controlled with the Nose-Hoover thermostat and barostat, respectively. The nominal pressure is maintained to 0~GPa. The Al-Sm sample is prepared by randomly substituting a certain fraction of Al atoms with Sm atoms in the solid state. The sample is first heated to 2000~K in order to melt it, then it is equilibrated at this high temperature for 300~ps, followed by a rapid cooling ($4 \times 10^{13}$~K/s) to 10~K to let the system vitrify. After an additional equilibration at 10~K, the as-quenched system is reheated quickly to the annealing temperature $T_\mathrm{anneal}$ for an isothermal nucleation reaction. The delay time $\tau$ for the first nucleation event since the start of the annealing process is recorded in every annealing simulation. An adaptive common neighbor analysis method~\cite{stukowski2012structure} was used for the structure identification, which examines the local environment to classify each atom as different structural types such as FCC, hexagonal close packed (HCP), body-centered cubic (BCC), or amorphous structure. Details of the temperature history, sample structural evolution, as well as the method to determine the delay time can be found in the Supplemental Materials.

In order to understand the Sm effect on nucleation, the nucleation data was fitted with the classical nucleation theory (CNT). From these fits the nucleation kinetic barriers were determined and then compared with Al/Sm diffusion energy barriers in order to identify the controlling process in the Al nucleation kinetics. Diffusion coefficients are calculated in the pre-nucleation state of the MG (see the Supplemental Materials). The details are reported in Secs.~\ref{sec_kinetics_controlled} and \ref{sec_mechanism}. Finally, the mechanisms participating in the nucleation event were identified; these mechanisms are compared to those governing Al diffusion in Al-Sm MG and are discussed in Sec.~\ref{sec_diffusion_attachment}.

One should note that, in this paper, the term ``nucleation barrier'' is avoided since it is ambiguous and does not distinguish the nucleation free energy barrier $\Delta G^*$ from the nucleation kinetic barrier $Q$. $\Delta G^*$ is the free energy cost associated with the formation of a critical nucleus of a new phase, whereas $Q$ is the energy barrier per atom that needs to be overcome in the process of atomic attachment to nuclei.

\section{Results}
\subsection{Time$-$temperature$-$transformation (TTT) curves}	
\begin{figure}[t]
	\centering
	\captionsetup[subfigure]{labelformat=empty}
	\begin{subfigure}{0.1\textwidth}
		\includegraphics[scale = 0.1]{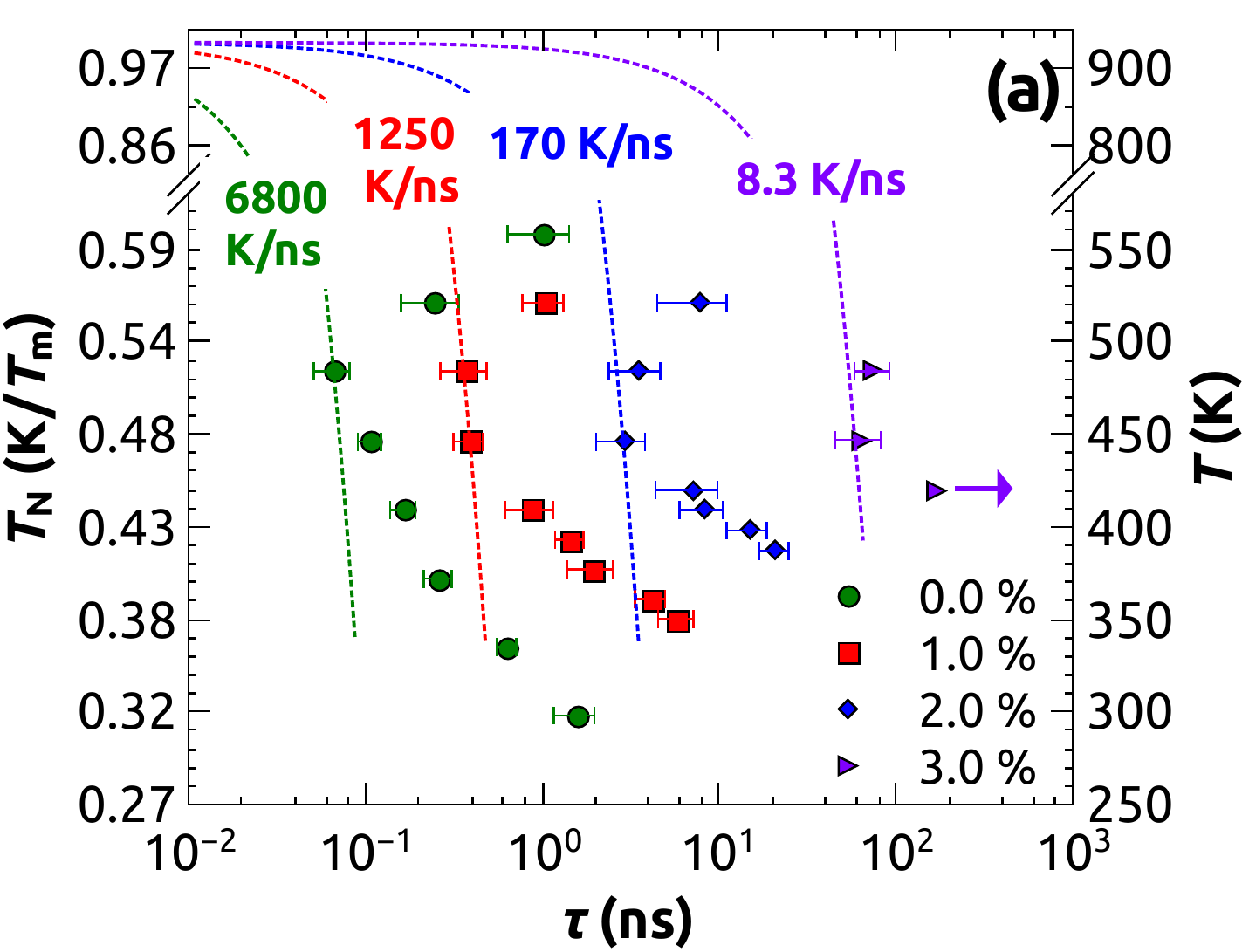} 
		\caption{\centering}
		\vspace{-5pt}
		\label{fig_ttt_curves}
	\end{subfigure}
	\thinspace\thinspace\enspace
	\begin{subfigure}{0.1\textwidth}
		\includegraphics[scale = 0.1]{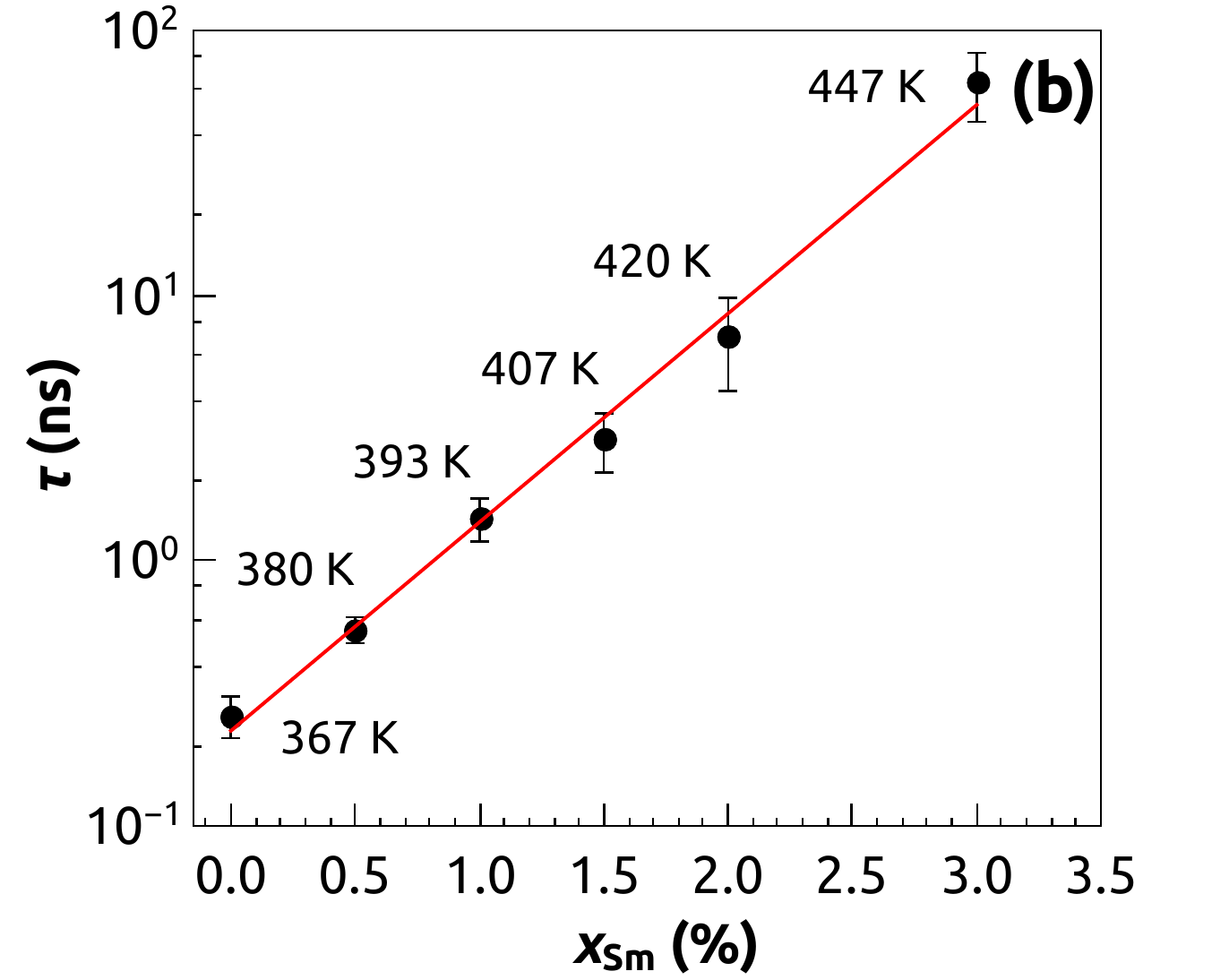}
		\caption{\centering}
		\vspace{-5pt}
		\label{fig_delay_time}
	\end{subfigure}
	\begin{subfigure}{0.1\textwidth}
		\includegraphics[scale = 0.1]{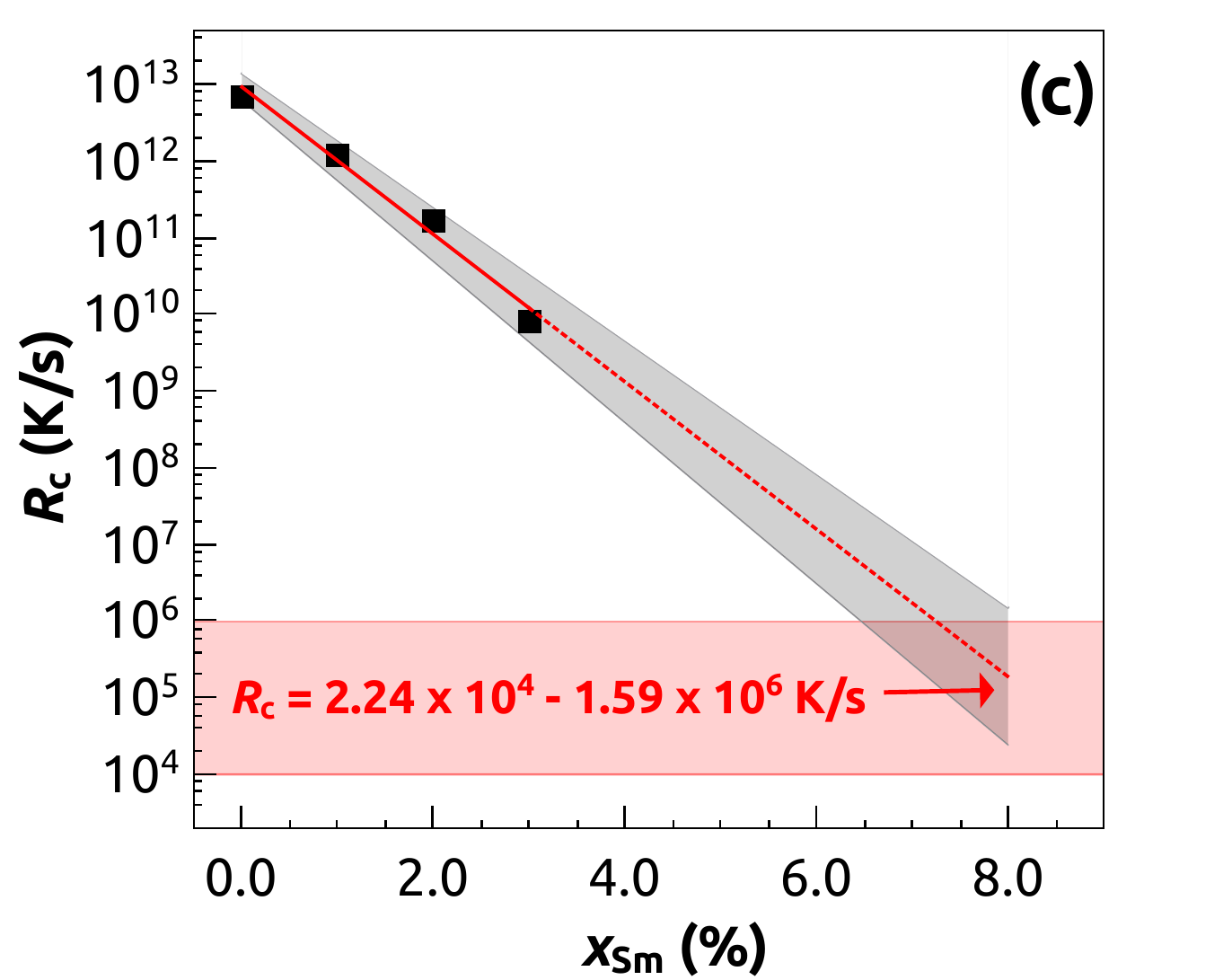}
		\caption{\centering}
		\vspace{-5pt}
		\label{fig_crit_cooling_rate}
	\end{subfigure}
	\vspace{-25pt}
	\caption{(Color online) (a) TTT curves for $x_\mathrm{Sm}$ = 0.0~at.\% (green circle), 1.0~at.\% (red rectangle), 2.0~at.\% (blue diamond) and 3.0~at.\% (violet right triangle). $T_\mathrm{N}$ represents the homologous temperature $T/T_\mathrm{m}$. The standard deviation is shown as the error bar. The continuous cooling curves are plotted as dashed lines and the values of $R_\mathrm{c}$ are indicated. (b) Delay time $\tau$ as a function of $x_\mathrm{Sm}$ at $0.95\,T_\mathrm{g}$. The solid line is the exponential fitting. (c) Critical cooling rate $R_\mathrm{c}$ as a function of $x_\mathrm{Sm}$. The data is fitted with the power relation and extrapolated to higher $x_\mathrm{Sm}$. The gray and red areas represent the fitting error bar and the reported range of $R_\mathrm{c}$ for marginal glass formers, respectively. The predicted range of $R_\mathrm{c}$ for $x_\mathrm{Sm} = 8.0~\mathrm{at}.\%$ is $2.2 \times 10^4$ - $1.6 \times 10^6$~K/s.}
	\label{fig_ttt_critCooling}
\end{figure}

Isothermal nucleation reactions were simulated for Al-Sm MGs with four Sm concentrations ($x_\mathrm{Sm} = $ 0.0, 1.0, 2.0 and 3.0~at.\%) at different $T_\mathrm{anneal}$. The time step was set to 10~fs for 3.0~at.\% and 2~fs otherwise. Ten independent simulations were run for each composition and each temperature. The average $\tau$ and the corresponding error bars are plotted in Fig.~\ref{fig_ttt_curves}. The measured delay times for different temperatures comprise the TTT curve that marks the onset of the crystallization transformation for each concentration. Fig.~\ref{fig_ttt_curves} shows that addition of Sm shifts the TTT curve towards the larger value of $\tau$ and therefore it retards the primary crystallization and enhances GFA of Al-Sm MGs. The ``nose'' temperatures ($T_\mathrm{nose}$) of the TTT curves are in the range of $0.48\,T_\mathrm{m}$ - $0.54\,T_\mathrm{m}$, where $T_\mathrm{m}$ is the melting temperature (933~K) of pure Al. The cooling process starts at $T_\mathrm{m}$ and the temperature decreases linearly with time $t$, i.e., $T = 933\,\mathrm{K} - R_\mathrm{c}\,t$. The corresponding continuous cooling curves tangential to the TTT curves at the ``nose'' temperature are also shown in Fig.~\ref{fig_ttt_curves}. Here, $R_\mathrm{c}$ is the estimated critical cooling rate. The estimated steady state nucleation rate $J_\mathrm{s}$ ($J_\mathrm{s} \approx e/(V\tau)$) was also calculated and it shows the opposite trend with temperature and Sm concentration (see Fig.~\ref{fig_nucleation_rate} in Supplemental Materials).

To determine quantitatively the effect of Sm on the nucleation delay time $\tau$, we anneal the systems at a fixed $T/T_\mathrm{g}$ ratio of 0.95. Here the glass transition temperature, $T_\mathrm{g}$, is calculated with the method described in Refs.~\cite{mendelev2015development} and \cite{mendelev2009development}, and the calculated values of $T_\mathrm{g}$ are shown in Table~\ref{tab_Tg}. The ratio of 0.95 is chosen because typical annealing experiments of Al-based MG use similar ratios~\cite{allen1998nanocrystal, perepezko2002amorphous}. $\tau$ as a function of $x_\mathrm{Sm}$ is plotted in Fig.~\ref{fig_delay_time}. Apart from the previously mentioned four Sm concentrations, simulations with 0.5~at.\% and 1.5~at.\% concentrations are also performed and reported here. The increasing trend of $\tau$ with $x_\mathrm{Sm}$ can be fitted with the power law: $\tau = C\!\cdot\!10^{k x_\mathrm{Sm}}$, where $C = 0.228$~ns and $k = 0.787$. The mechanism underlying this significant effect of Sm on $\tau$ will be discussed in Secs.~\ref{sec_kinetics_controlled} and \ref{sec_mechanism}. The fact that the error bar of $\tau$ (note the logarithmic scale) increases with $x_\mathrm{Sm}$ implies a more stochastic nature of the nucleation process at higher Sm concentrations.

\begin{table}[h]
	\centering
	\vspace{-4pt}
	\setlength{\tabcolsep}{15pt}
	\begin{tabular}{ccc}
		\toprule
		$x_\mathrm{Sm}$(at.\%)	& $T_\mathrm{g}$(K)	& $T_\mathrm{anneal}$(K)	\\\midrule
		0.0						& 386					& 367						\\
		0.5						& 400					& 380						\\
		1.0						& 414					& 393						\\
		1.5						& 428					& 407						\\
		2.0						& 442					& 420						\\
		3.0						& 471					& 447						\\
		\bottomrule
	\end{tabular}
	\caption{Calculated $T_\mathrm{g}$ and $T_\mathrm{anneal}$ ($T_\mathrm{anneal} = 0.95\,T_\mathrm{g}$).}
	\label{tab_Tg}
\end{table}

The critical cooling rate $R_\mathrm{c}$ as a function of $x_\mathrm{Sm}$ is plotted in Fig.~\ref{fig_crit_cooling_rate}. It is found that the decay of $R_\mathrm{c}$ with increasing $x_\mathrm{Sm}$ is well approximated by a power function, i.e., $R_\mathrm{c} = C^\prime\!\cdot\!10^{-k^\prime x_\mathrm{Sm}}$, where $C^\prime = (9.9\pm 3.8)\times 10^{12}$~K/s and $k^\prime = 0.96 \pm 0.09$. Our predictions are in a good agreement with available literature data. For instance, MD simulations by Hou~\textit{et al.}~\cite{hou2015formation} with an embedded atom method interatomic potential~\cite{mendelev2008analysis} have shown that $R_\mathrm{c}$ for pure Al is within the range of $4.0 \times 10^{12}$ - $1.0 \times 10^{13}$~K/s. This result is consistent with that of $6.8 \times 10^{12}$~K/s found in our work. On the other hand, if the power relation determined in our simulations is extrapolated to higher $x_\mathrm{Sm}$, $R_\mathrm{c}$ is predicted to be $2.24 \times 10^4$ - $1.59 \times 10^6$~K/s for $x_\mathrm{Sm} = 8.0~\mathrm{at}.\%$, which corresponds to the composition of a well-known marginal glass former~\cite{foley1996analysis, wilde1999glass}. This predicted range is consistent with experimental data~\cite{senkov2001effect, suzuki1983formation}, which showed that general Al-based MGs have $R_\mathrm{c}$ in the range of $10^4$ - $10^6$~K/s.

\subsection{Nucleation kinetics} \label{sec_kinetics_controlled}
In order to understand the Sm effect on the nucleation kinetics, our simulation data was evaluated in the light of the CNT for homogeneous nucleation. From CNT, the steady-state nucleation rate $J_\mathrm{s}$, defined as the number of nuclei in unit volume per unit time, can be written as~\cite{walton1969nucleation, markov2003crystal}

\begin{equation}
	J_\mathrm{s} = \omega^* Z N^*, \label{eq_nucleation}
\end{equation}
where $\omega^*$, $Z$ and $N^*$ are the attachment frequency of monomers to the critical nucleus, the Zeldovich factor, and the equilibrium concentration of critical nuclei, respectively. $\omega^*$ and $N^*$ represent the kinetic and thermodynamic contributions, respectively, to the overall nucleation rate, and they can be calculated as
\begin{equation}
	\omega^* = A C \nu \lambda \exp\left(-\frac{Q}{k_\mathrm{B}T}\right), \label{eq_kinetic_controlled}
\end{equation}
\begin{equation}
	N^* = N_1 \exp\left(-\frac{\Delta G^*}{k_\mathrm{B}T}\right). \label{eq_driving_controlled}
\end{equation}
Here, $A$ is the nucleus surface area, $C$ is the concentration expressed in number of monomers per volume, $\nu$ is a frequency factor, $\lambda$ is the mean free path of Al in MG, $N_1$ is Al monomer concentration, $T$ is temperature, $k_\mathrm{B}$ is the Boltzmann constant, $Q$ is the nucleation kinetic barrier, and $\Delta G^*$ is the free energy barrier determined by nucleation driving force $\Delta G_\mathrm{v}$ and interfacial energy $\gamma$~\cite{markov2003crystal}. For spherical nuclei, $\Delta G^* = 16\pi\gamma^3/(3\Delta G_\mathrm{v}^2)$~\cite{abraham1974homogeneous}.

Nucleation is controlled by thermodynamics at shallow supercoolings ($T > T_\mathrm{nose}$) due to the smaller nucleation driving force and it is controlled by kinetics at deep supercoolings ($T < T_\mathrm{nose}$) due to the slower atomic mobility. The balance of these thermodynamic and kinetic factors results in the highest nucleation rate at $T_\mathrm{nose}$. In our simulations a number of sub-critical Al FCC clusters have been observed at deep supercoolings, and their formation is due to the high nucleation driving force. However, these sub-critical clusters grow very slowly and very few of them have a chance to grow to the super-critical size, indicating a sluggish nucleus attachment kinetics and resulting in the prolonged delay time at deep supercoolings. Therefore, at $T < T_\mathrm{nose}$, the term $\exp(-Q/k_\mathrm{B}T)$ dominates, and Eq.~\eqref{eq_nucleation} can be rewritten as $J_\mathrm{s} = K \exp(-Q/k_\mathrm{B}T)$ with $K$ as approximately constant. Since the delay time $\tau$ is inversely proportional to $J_\mathrm{s}$~\cite{dantzig2009solidification}, therefore, one can write $\tau \propto \exp(Q/k_\mathrm{B}T)$. 

\begin{figure}[h]
	\centering
	\includegraphics[scale = 0.6]{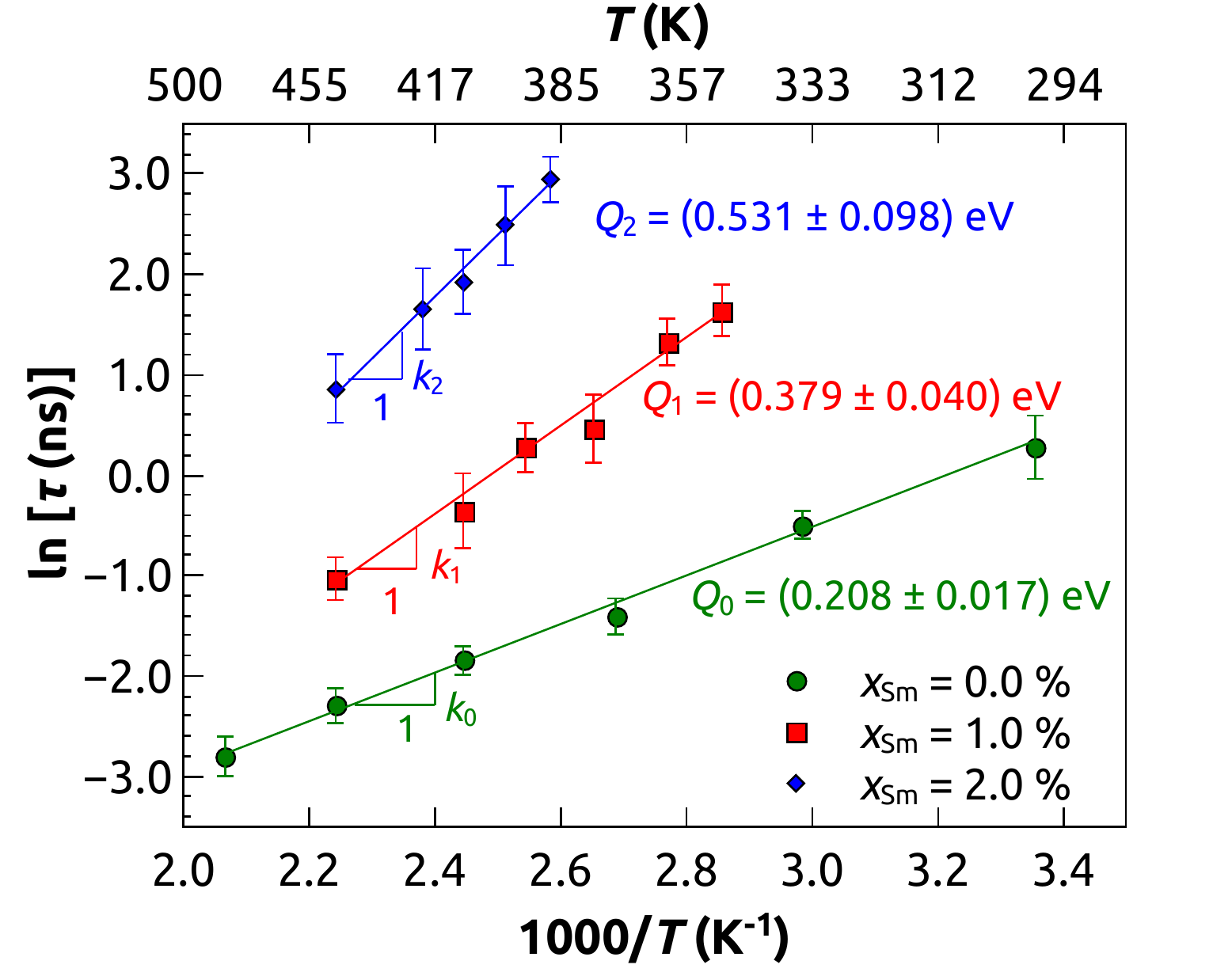}
	\caption{(Color online) $\ln (\tau)$ vs. $1/T$ for $T < T_\mathrm{nose}$. The solid lines are the weighted exponential fitting of delay time $\tau$. The nucleation kinetic barrier $Q_\mathrm{i}$ is calculated as $Q_\mathrm{i} = k_\mathrm{i}\,k_\mathrm{B}$, where $k_\mathrm{i}$ represents the slope of the fitting line.}
	\label{fig_kinetic_barrier}
\end{figure}

Fig.~\ref{fig_kinetic_barrier} plots $\ln (\tau)$ vs. $1/T$  in the temperature range of $T < T_\mathrm{nose}$, along with the fits. The nucleation kinetic barrier $Q_\mathrm{i}$ is calculated as $Q_\mathrm{i} = k_\mathrm{i}\,k_\mathrm{B}$, where $k_\mathrm{i}$ is the slope of the fitting line. The values of $Q$ for $x_\mathrm{Sm} =$ 0.0, 1.0 and 2.0~at.\% are $0.208 \pm 0.017$, $0.379 \pm 0.040$ and $0.531 \pm 0.098$ eV, respectively. Note that $Q$ for 2.0~at.\% Sm concentration has a large error bar, because the long nucleation delay time restricts the temperature range that can be studied for this concentration. Clearly, Sm additions raise the nucleation kinetic barrier $Q$ and therefore they significantly reduce the Al attachment rate $\omega^*$ to nuclei. As a result, the nucleation delay time increases exponentially with increasing Sm as shown in Fig.~\ref{fig_delay_time}.

\subsection{Rate-limiting process in nucleation} \label{sec_mechanism} 
Having shown that Sm solutes remarkably reduce the frequency of Al attachment to nuclei, an interesting question that arises is what is the controlling process in Al nucleation kinetics. It has been shown that the atomic transport in nucleation can be governed either by bulk diffusion or by interface attachment, depending on specific circumstances~\cite{kashchiev2000nucleation, kashchiev2003review, duran2015unification}.

\begin{figure}[t]
	\centering
	\includegraphics[scale = 0.65]{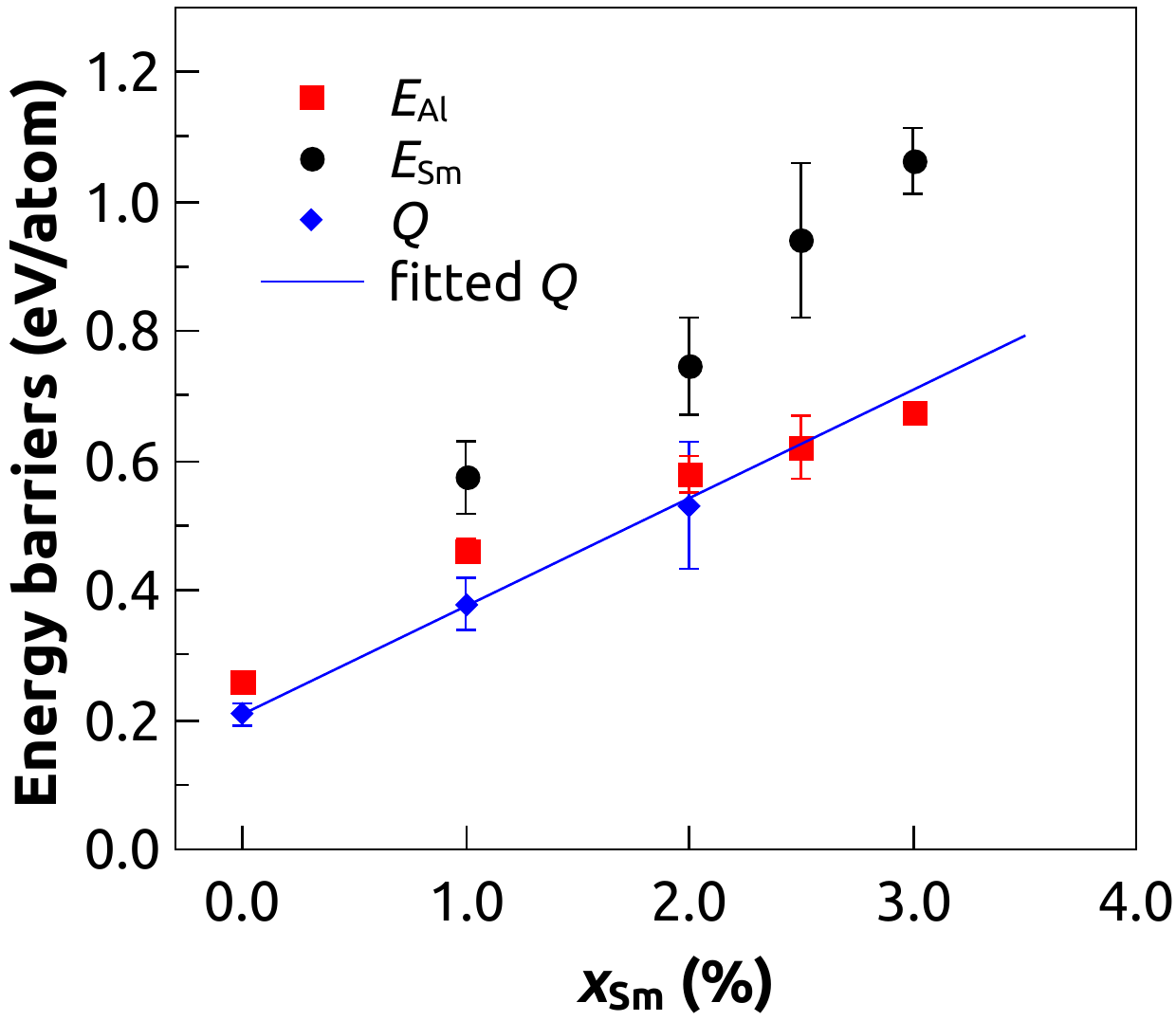}
	\caption{(Color online) Nucleation kinetic barrier ($Q$) and diffusion energy barriers ($E_\mathrm{Al}$ and $E_\mathrm{Sm}$). The sizes of the error bars represent standard deviations in the data.}
	\label{fig_comparison_kinetic_barrier_diffusion_activation}
\end{figure}

In order to discover the process controlling the development of Al nucleus, we calculate the diffusion energy barriers ($E_\mathrm{Al}$ and $E_\mathrm{Sm}$ for Al and Sm, respectively) and compare them with the nucleation attachment barrier $Q$. $E_\mathrm{Al}$ and $E_\mathrm{Sm}$ are obtained by fitting the diffusion coefficients determined during annealing before any nucleation took place. Details on the calculation of diffusion coefficients and diffusion energy barriers can be found in the Supplemental Materials. The nucleation kinetic barrier ($Q$) and the diffusion energy barriers ($E_\mathrm{Al}$ and $E_\mathrm{Sm}$) are plotted together in Fig.~\ref{fig_comparison_kinetic_barrier_diffusion_activation} for comparison. The fact that the nucleation kinetic barrier $Q$ is smaller than $E_\mathrm{Sm}$ indicates that the attachment frequency is controlled by a process that is faster than Sm bulk diffusion. There is a relatively good overall agreement between $Q$ and $E_\mathrm{Al}$, which in principle might mean that Al diffusion is the rate controlling process. However, this is not a valid explanation because the Sm concentration is very low and the prevalence of Al atoms near nuclei makes the long-distance diffusion of Al unnecessary for nucleation. This analysis implies that the attachment frequency is interface-controlled and Al atomic transfer across the interface (constituting an ``attachment'' event) is the rate limiting process in the composition range considered in this study. 

It is interesting that the interface attachment of Al to a nucleus would have a comparable energy barrier to the barrier for Al diffusion through the bulk MG. In order to shed light on this observation, the specific mechanisms underlying both Al diffusion and attachment will be examined and compared to each other in the following section.

\subsection{Diffusion and nucleation attachment mechanism} \label{sec_diffusion_attachment}
Unlike crystalline metals where diffusion often occurs by means of single-atom jumps via a vacancy or interstitial mechanism, MD simulations have shown that diffusion in many MGs takes place in a heterogeneous manner (i.e., in a given time interval there are diffusing and non-diffusing regions of the structure) and that diffusion occurs through collective movement of a chain or ring of atoms~\cite{faupel2003diffusion, zhang2015diffusion}. Here, first a quantitative analysis was conducted for the simulated samples to confirm that Al diffusion in our system follows the same collective pattern as reported in many other MGs~\cite{faupel2003diffusion, zhang2015diffusion}. Then it will be shown that during nucleation, attachment of Al atoms to nuclei is also heterogeneous and involves a collective motion of atoms.

\subsubsection{Al diffusion mechanism}
\label{sec_al_diffusion}
To elucidate the diffusion mechanism in MGs, it is first necessary to identify the diffusing atoms (DAs) and the diffusion events. Here, the DAs are defined as those which perform displacements larger than a cut-off $d_\mathrm{c}$ within a time window $\Delta t$. All the DAs identified in this section are Al atoms, as Sm atoms displace too slowly and are not diffusing by the cut-off definition. In fact, the diffusion coefficient of Sm atoms is at least one order of magnitude lower than that of Al, as shown in Fig.~\ref{fig_diffusion_activation} in Supplemental Materials. The fraction of DAs among all the atoms in the sample is usually smaller than a few percent and it decreases with increasing $d_\mathrm{c}$ and decreasing $\Delta t$. The number and fraction of DAs for different $d_\mathrm{c}$ and $\Delta t$ can be found in Fig.~\ref{fig_Num_DiffAl} in the Supplemental Materials. Once the DAs have been identified, the spatial correlations among these atoms are analyzed to address the question of whether Al diffusion is heterogeneous. For this purpose, the nearest neighbor analysis (NNA) and reduced number density analysis are employed on all the DAs. 

\begin{figure}[t]
	\centering
	\captionsetup[subfigure]{labelformat=empty}
	\begin{subfigure}{0.34\textwidth}
		\centering
		\includegraphics[scale = 0.425]{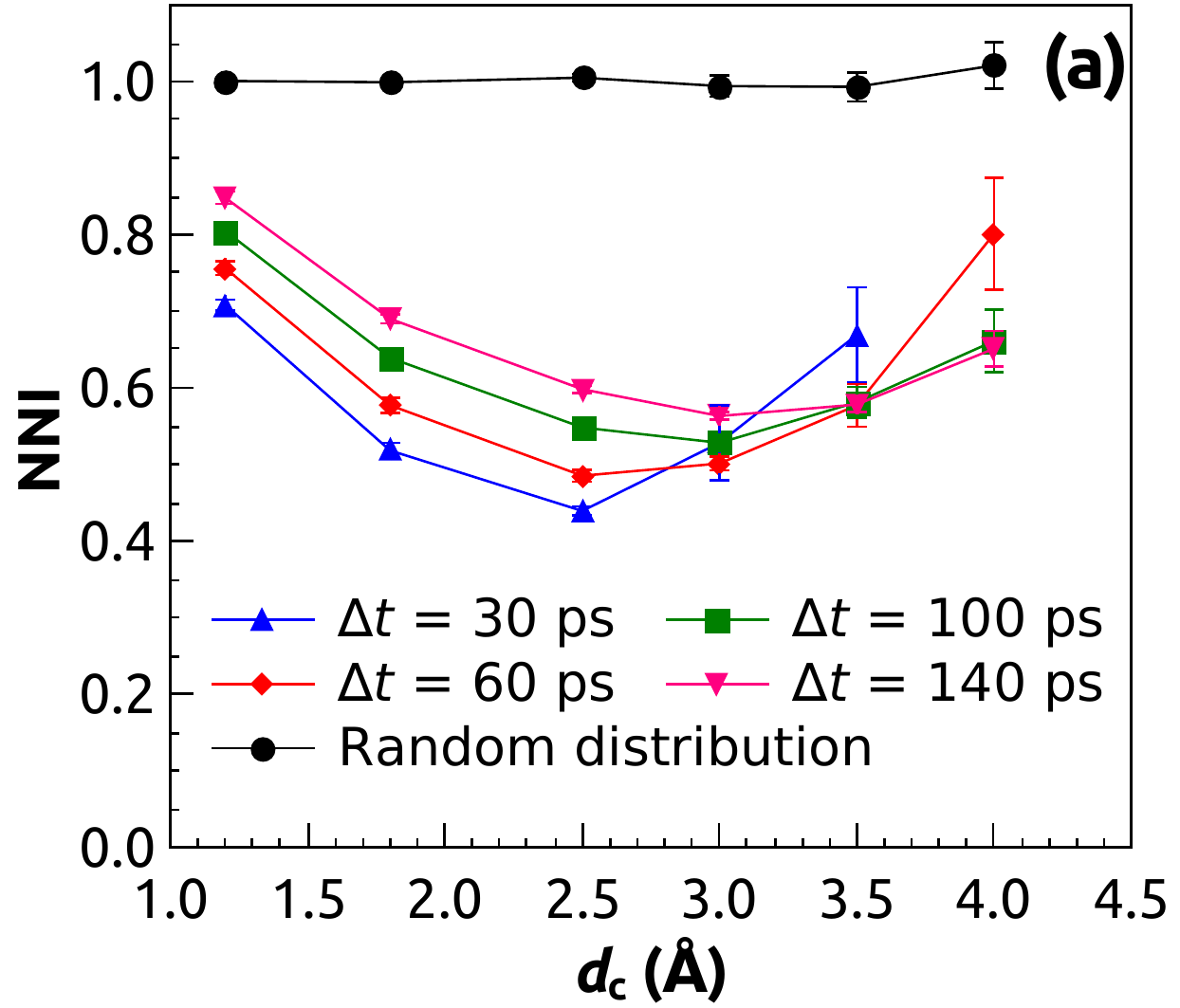}
		\caption{\centering}
		\vspace{-5pt}
		\label{fig_NNI}
	\end{subfigure}
		\begin{subfigure}{0.30\textwidth}
		\centering
		\includegraphics[scale = 0.415]{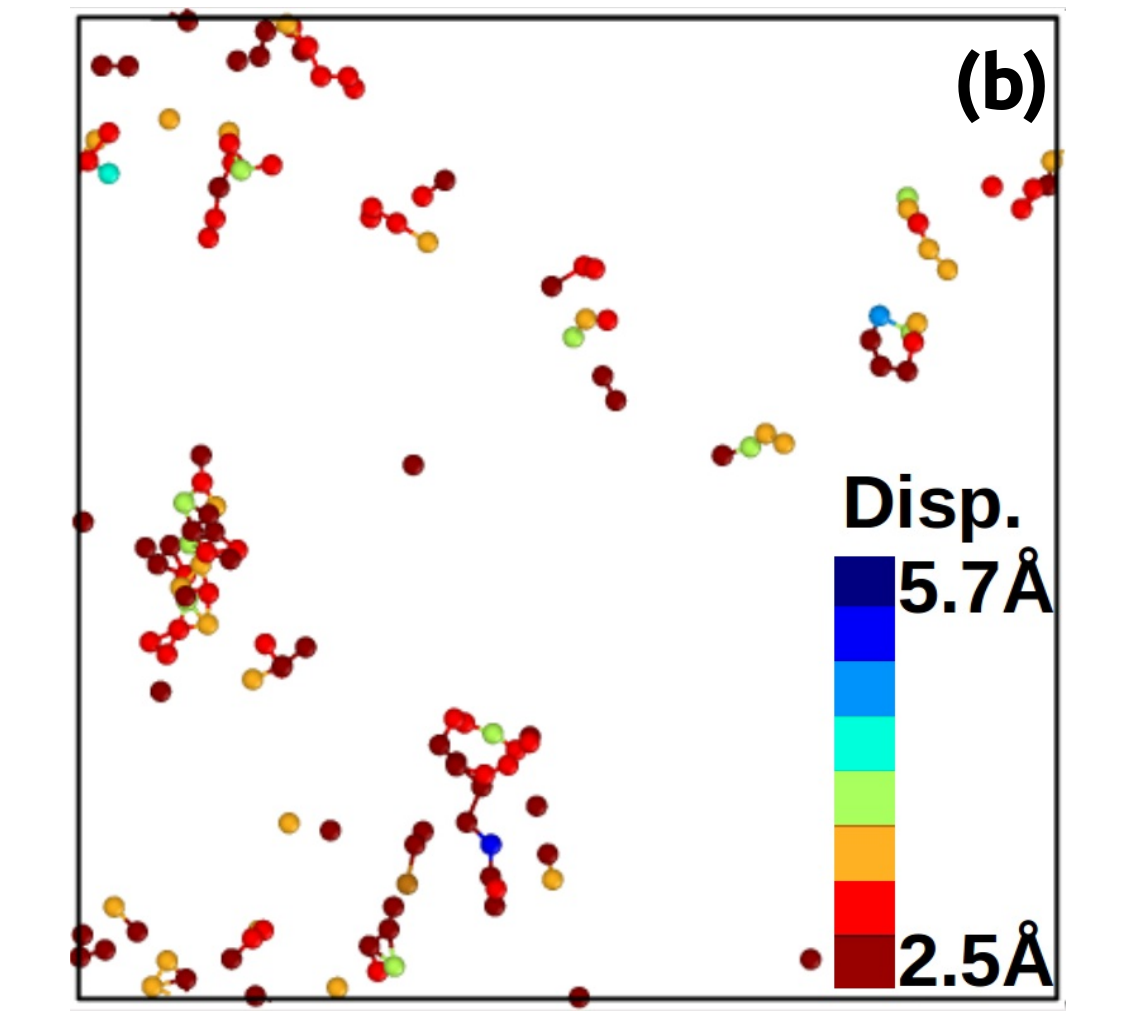}
		\caption{\centering}
		\vspace{5pt}
		\label{fig_collective_diffAl}
	\end{subfigure}
	\begin{subfigure}{0.34\textwidth}
		\centering
		\includegraphics[scale = 0.425]{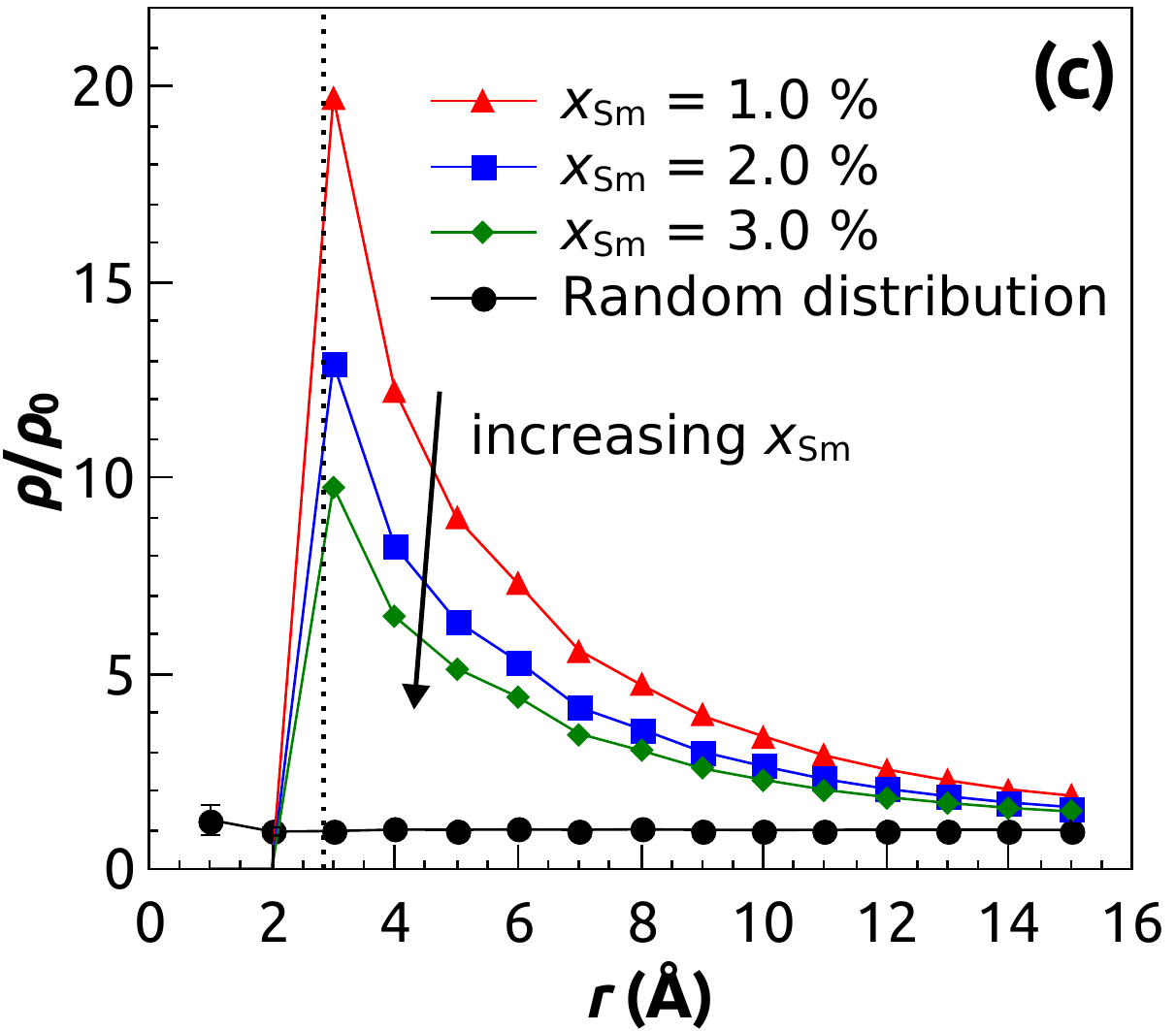}
		\caption{\centering}
		\vspace{-5pt}
		\label{fig_num_density}
	\end{subfigure}
	\vspace{-25pt}
	\caption{(Color online) (a) NNI of DAs as a function of displacement cut-off $d_\mathrm{c}$ and time window $\Delta t$. (b) A 1-nm-thick 2D slice of the DAs for $d_\mathrm{c} = 2.5$~\AA~ and $\Delta t = 100$~ps. The color represents displacement magnitude. (c) Reduced number density of DAs around the central DA. Here $d_\mathrm{c} = 2.5$~\AA~ and $\Delta t = 60$~ps are used to define diffusing events. The vertical dashed line represents the nearest Al-Al distance (2.8~\AA) in the system. (a) and (b) are obtained from the system with $x_\mathrm{Sm}$ = 2.0~at.\%. For comparison, in (a) and (c), respectively, we also plot NNI and the reduced number density for randomly distributed atoms with the same number of atoms as DAs in our Al-Sm samples. In (c) the data corresponds to $\Delta t = 60$~ps.}
	\label{fig_diffusion_collective}
\end{figure}

In NNA, the average distance $d_\mathrm{nn}$ between the nearest neighboring DAs is first calculated. The nearest neighbor index (NNI) is then defined as $d_\mathrm{nn}/d_\mathrm{ran}$, where $d_\mathrm{ran} = 0.55396\,N_\mathrm{v}^{-\frac{1}{3}}$ is the theoretical average nearest neighbor distance for the same number of randomly distributed atoms~\cite{chandrasekhar1943stochastic}. From NNI one can learn whether the DAs are clustered (NNI~$< 1$), randomly distributed (NNI~$\approx 1$), or more uniformly distributed than random (NNI~$> 1$). Fig.~\ref{fig_NNI} shows an example of a calculated NNI for a relaxed system with $x_\mathrm{Sm}$ = 2.0~at.\% and $T = 0.95\,T_\mathrm{g}$ before nucleation occurs. Since NNI~$< 1$, this means the DAs tend to cluster and they perform diffusing displacements collectively. Different choices of $\Delta t$ lead to qualitatively the same results. Fig.~\ref{fig_collective_diffAl} presents a 2D view of DAs for $d_\mathrm{c} = 2.5$~\AA~ and $\Delta t = 100$~ps. This analysis shows that the distribution of DAs is highly heterogeneous and that DAs form chains or rings, confirming the cooperative nature of Al diffusion.

In addition to NNA, the reduced number density $\rho/\rho_0$ of DAs was also calculated. Here, $\rho$ is the number density of DAs around the central DA and is calculated as $\rho = N(r)/(\frac{4}{3}\pi r^3)$, where $N(r)$ is the DA count within a distance $r$ from the central DA. $\rho_0$ is the average number density of DAs in the system. Fig.~\ref{fig_num_density} shows $\rho/\rho_0$ of DAs for different Sm concentrations at $T = 0.95\,T_\mathrm{g}$ using $d_\mathrm{c} = 2.5$~\AA~ and $\Delta t = 60$~ps. It is evident that $\rho/\rho_0$ increases with decreasing $r$ and reaches a maximum value at around the nearest Al-Al distance, which indicates that the DAs are prone to become direct nearest neighbors of each other. This is consistent with the observation in Fig.~\ref{fig_collective_diffAl}. The decrease of $\rho/\rho_0$ with increasing Sm concentration suggests the decreasing size of the DA-cluster. This trend is attributed to the reduced size of the liquidlike diffusion channels with increasing Sm concentration, as discussed later in this section.

\begin{figure}[t]
	\centering
	\captionsetup[subfigure]{labelformat=empty}
	\begin{subfigure}{0.495\textwidth}
		\centering
		\includegraphics[scale = 0.59]{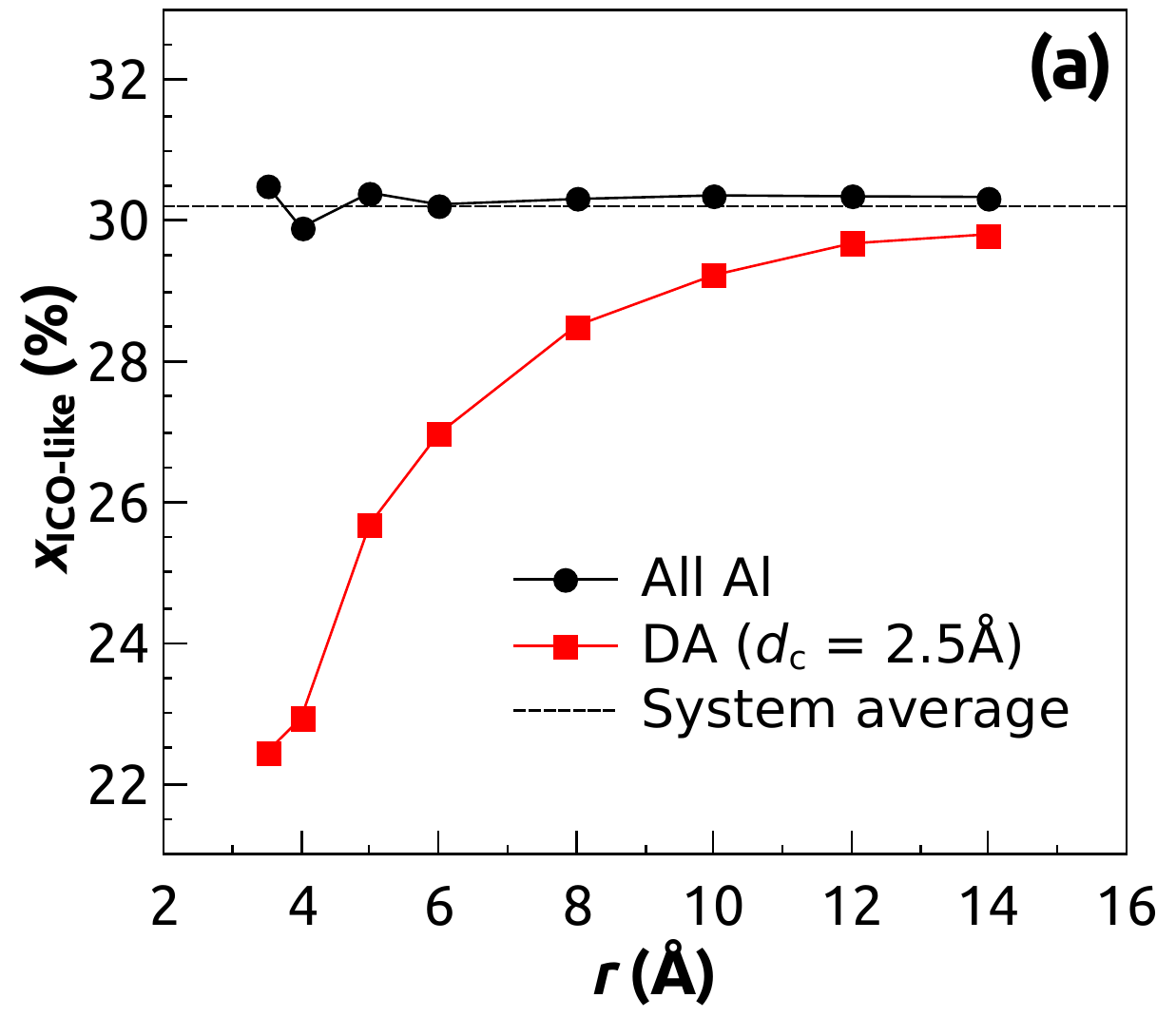}
		\caption{\centering}
		\vspace{-5pt}
		\label{fig_diffusion_avoidICOlike}
	\end{subfigure}
	\begin{subfigure}{0.495\textwidth}
		\centering
		\includegraphics[scale = 0.59]{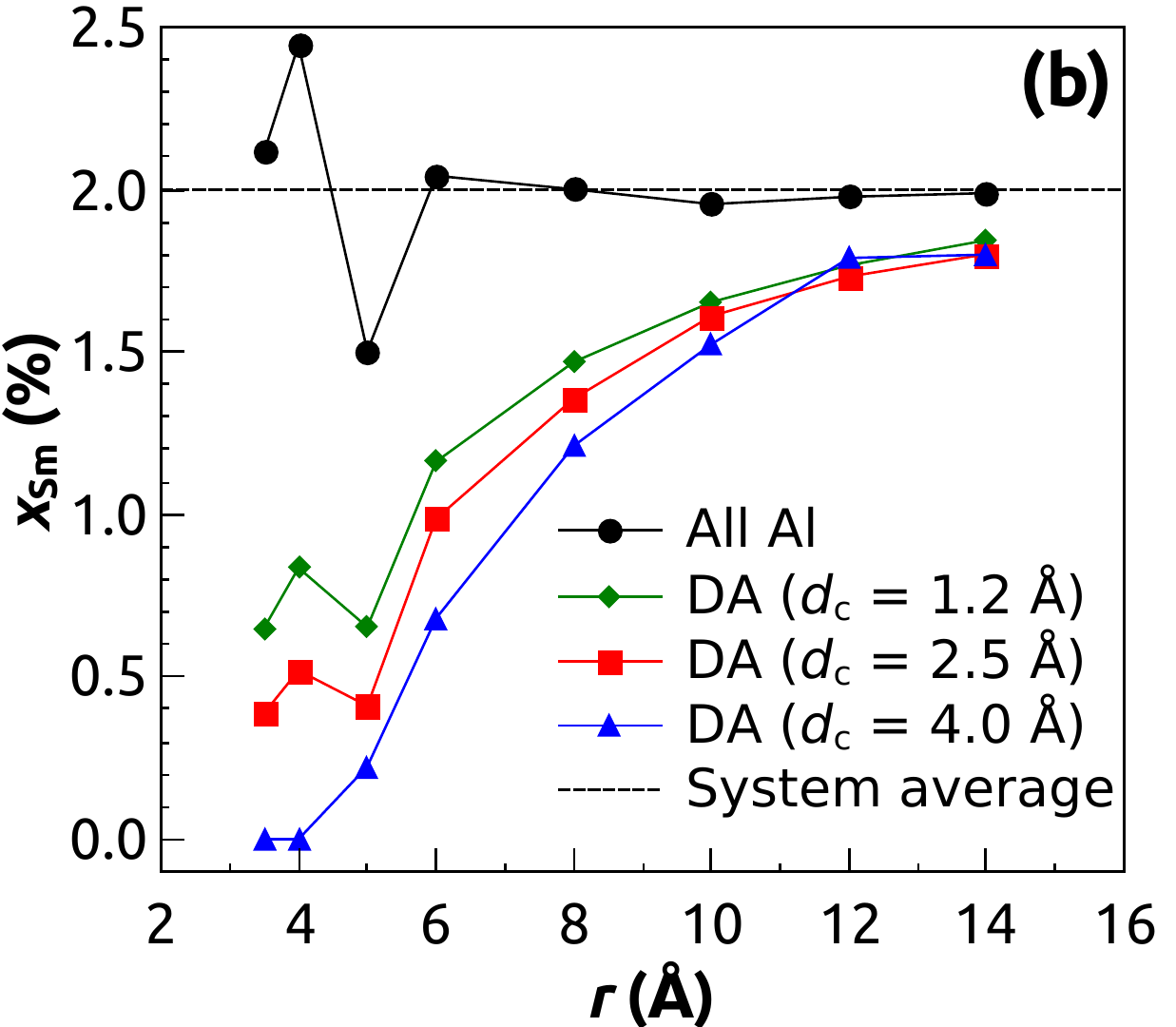}
		\caption{\centering}
		\vspace{-5pt}
		\label{fig_diffusion_avoidSm}
	\end{subfigure}
	\vspace{-25pt}
	\caption{(Color online) (a) ICO-like VP concentration $x\textsubscript{ICO-like}$ and (b) Sm concentration $x_\mathrm{Sm}$ around the central diffusing Al atom (DA). These concentrations around any Al atom and the system average concentrations are also plotted for comparison. The diffusing events are defined using $\Delta t = 100$~ps and different $d_\mathrm{c}$.}
	\label{fig_diffusion_avoidICOSm}
\end{figure}

To elucidate further the collective nature of DAs, their spatial correlations are analyzed with the icosahedral Voronoi polyhedra (ICO VP) and Sm atoms. The motivation for this analysis is that the concentration of ICO VPs has been previously argued to be correlated with the GFA of an alloy~\cite{cheng2008alloying, peng2010effect}. In this case, the ICO VPs are identified by applying the Voronoi tessellation technique~\cite{finney1970random} and then grouped based on their Voronoi indices~\cite{bokas2016role}. The concentrations of the center atoms of ICO-like VPs and Sm atoms within a distance $r$ from any DA are calculated and plotted in \cref{fig_diffusion_avoidICOlike,fig_diffusion_avoidSm}, respectively. According to Fig.~\ref{fig_diffusion_avoidICOlike} there are fewer ICO-like VPs near DAs, implying that DAs tend to avoid ICO VPs. This observation is consistent with findings previously reported in literature. Specifically, Zhang~\textit{et al.}~\cite{zhang2015diffusion} reported that the diffusion of Cu in a Cu-Zr MG is confined in the liquidlike regions that are poor in icosahedral short-range order, and Bokas~\textit{et al.}~\cite{bokas2016role} found that in Al-Sm MGs the diffusion coefficient of Al with ICO-like VP is one order of magnitude lower than other Al atoms at low Sm concentration range. In the same vein, Fig.~\ref{fig_diffusion_avoidSm} shows that DAs are prone to avoid Sm atoms in the sample and that faster DAs (i.e., DAs that have larger displacements within a fixed time window) have stronger avoidance tendency. We speculate that Sm can have an important effect on Al diffusion by increasing the fraction of ICO-like VPs~\cite{bokas2016role}. This would take place not only because Sm increases the fraction of Al-centered ICO-like VPs, but also because most of the Sm-centered clusters are ICO-like. As a result, the average size of liquidlike channels is reduced and it takes a longer time for Al atoms to navigate the torturous diffusion paths inside the narrower channels. Finally, the fact that the sizes of liquidlike channels and DA-clusters correlate with each other explains the decreasing size of DA-clusters with increasing $x_\mathrm{Sm}$, as shown in Fig.~\ref{fig_num_density}.

\subsubsection{Nucleation attachment mechanism}
\label{sec_nucleation_attachment}

\begin{figure}[h]
	\centering
	\includegraphics[scale = 0.25]{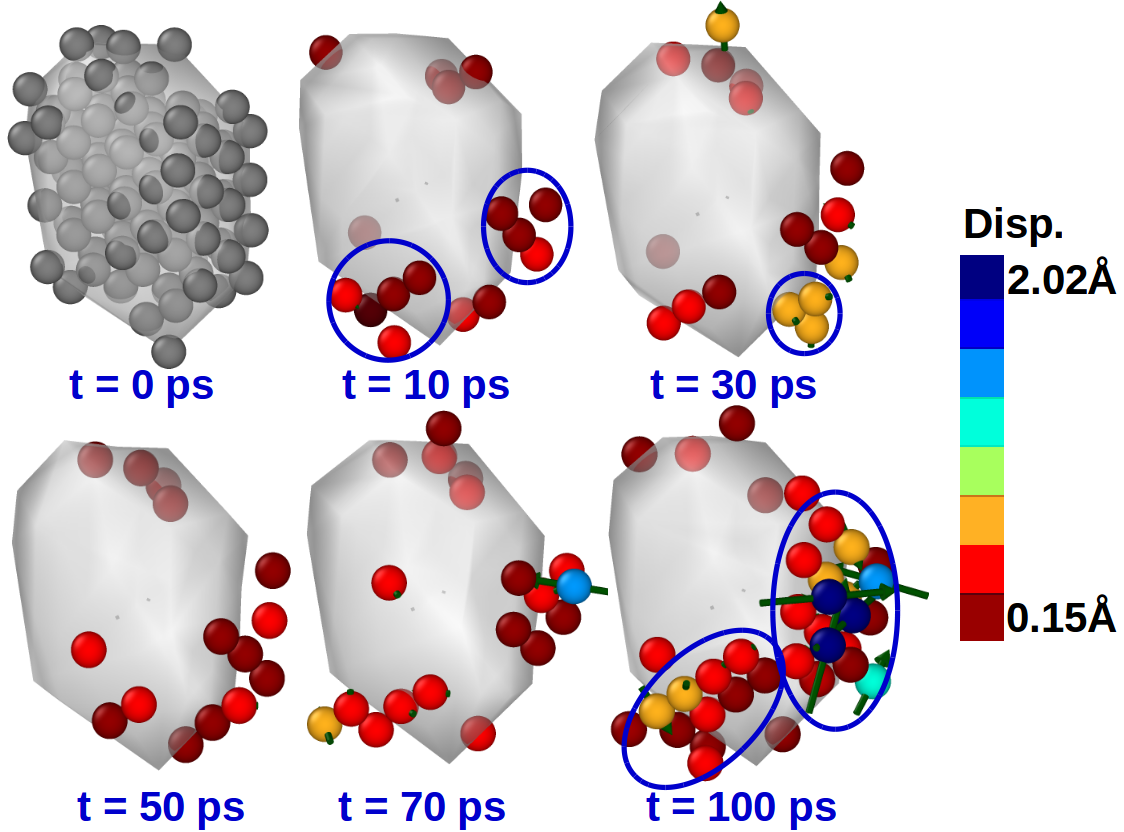}
	\vspace{-2pt}
	\caption{(Color online) Nucleus attachment process for $x_\mathrm{Sm}$ = 2.0~at.\% and $T = 0.95\,T_\mathrm{g}$. The isosurface represents the original nucleus surface at $t = 0$~ps. The color atoms are those that attach to nucleus subsequently. Color of the atoms and arrow represent the magnitude and direction of the displacement vector of a given atom.}
	\label{fig_attachment_collective}
\end{figure}

In order to illustrate the mechanism of Al attachment during nucleation, Fig.~\ref{fig_attachment_collective} presents the snapshots of Al attachment process during the growth of a nucleus in the system with $x_\mathrm{Sm}$ = 2.0~at.\% and $T = 0.95\,T_\mathrm{g}$. In the first image, the grey atoms are the original FCC atoms at $t = 0$~ps and they are used to construct an isosurface shown also in grey. In the following images (at longer times), only newly attached atoms are shown explicitly, and the colors and arrows, respectively, represent the magnitude and the direction of the atom's displacement relative to its position at $t = 0$~ps. From the images shown in Fig.~\ref{fig_attachment_collective}, one can see that the attachment is heterogeneous in space and atoms tend to attach to nuclei collectively as a cluster. Examples of these clusters are marked with the blue ellipses in the images in Fig.~\ref{fig_attachment_collective}. 

\begin{figure}[h]
	\centering
	\includegraphics[scale = 0.6]{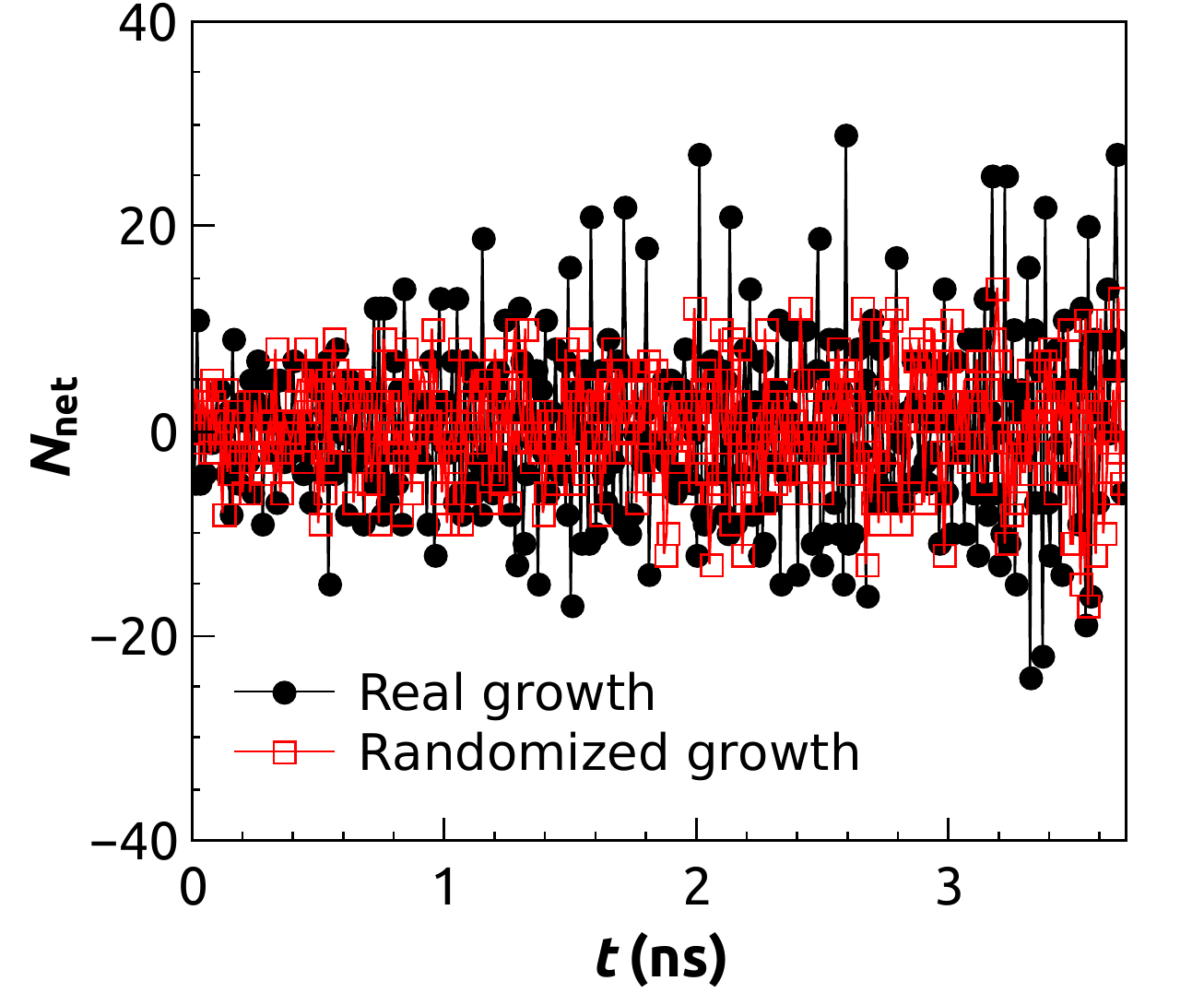}
	\caption{(Color online) The instantaneous net increase, $N_\mathrm{net}$, of the number of atoms in the nucleus within 10~ps for real and randomized nucleus growth.}
	\label{fig_attachment_heter_jump}
\end{figure}

The attachment is not only heterogeneous in space, but also in time. The temporal heterogeneity of attachment was verified by tracking the dynamic fluctuation of the instantaneous net increase $N_\mathrm{net}$ of the number of atoms added to the nucleus within the period of 10~ps, i.e., $N_\mathrm{net} = N_\mathrm{att} - N_\mathrm{det}$, where $N_\mathrm{att}$ and $N_\mathrm{det}$ are the instantaneous numbers of attachment and detachment events within 10~ps, respectively. Fig.~\ref{fig_attachment_heter_jump} shows $N_\mathrm{net}$ vs. time $t$ for a nucleus growing from 17 atoms to 189 atoms within 3.7~ns; during this time there are in total 4757 attachment events and 4585 detachment events. One can see that $N_\mathrm{net}$ exhibits strong fluctuations, which indicates the temporal heterogeneity of the nucleus growth. To further demonstrate this temporal heterogeneity, the simulation method was modified to numerically model a randomized nucleus growth. For the analysis, the following simplifying assumptions were made: 1) The nucleus is spherical and the nucleus surface area $A = (4\pi)^\frac{1}{3} (3NV_0)^\frac{2}{3}$, where $N$ and $V_0$ are the number of nucleus atoms and the atomic volume, respectively; 2) the probability of attachment and detachment at any time is proportional to the nucleus surface area; and 3) the numbers of attachment and detachment events are the same as found in the full MD simulations with $x_\mathrm{Sm}$ = 2.0~at.\% and $T = 0.95\,T_\mathrm{g}$, i.e., $N_\mathrm{att} = 4757$ and $N_\mathrm{det} = 4585$. The randomized fluctuation of $N_\mathrm{net}$ based on single-atom attachment/detachment is also plotted in Fig.~\ref{fig_attachment_heter_jump}. Clearly, the $N_\mathrm{net}$ fluctuation amplitude in randomized nucleus growth is much smaller than in the case of the real nucleus growth. This means that attachment of atoms to nuclei in a real system is not random and it is heterogeneous in time. The difference in $N_\mathrm{net}$ fluctuation in Fig.~\ref{fig_attachment_heter_jump} can be attributed to the different nucleus growth modes (i.e., single atom attachment versus collective attachment) in these two systems. 

In summary, our analysis shows that both Al diffusion and Al nucleation attachment occur highly heterogeneously via collective movements of a group of atoms. The common features of these physical processes may explain why the kinetic barrier $Q$ of Al nucleation is comparable to the activation barrier $E_\mathrm{Al}$ of Al diffusion, as shown in Fig.~\ref{fig_comparison_kinetic_barrier_diffusion_activation}.

\section{Discussion}
It has been suggested~\cite{russell1980nucleation} that the homogeneous nucleation delay time $\tau$ is inversely proportional to the atomic attachment rate to nucleus, that is, $\tau \propto 1/\omega^*$. Combining the power law of $\tau$ (see Fig.~\ref{fig_delay_time}) and Eq.~\eqref{eq_kinetic_controlled}, taking the logarithm of both sides, and omitting constant terms, one can find that $Q \propto x_\mathrm{Sm}$. This analysis is consistent with data shown in Fig.~\ref{fig_comparison_kinetic_barrier_diffusion_activation}, where the dependence of the mean value of the attachment kinetic barrier $Q$ on Sm concentration $x_\mathrm{Sm}$ can be well approximated by a linear function (despite the large error bar of $Q$ at higher concentrations). This consistence implies a good agreement between our simulations and the theory.

Our simulation results show that, in the composition range (0.0 - 3.0~at.\% Sm) studied, the nucleation kinetics in Al-Sm MG is governed by the interfacial attachment process. Mechanisms underlying Al interfacial attachment and Al diffusion share critical atomic-level features, i.e., they both occur heterogeneously through collective movements of a group of atoms. The fact that the nucleation kinetic barrier $Q$ is smaller than $E_\mathrm{Sm}$, as seen in Fig.~\ref{fig_comparison_kinetic_barrier_diffusion_activation}, suggests that the bulk diffusion of Sm is not involved in the nucleation process. As the Sm concentration is low, most Sm atoms might be transported away from the nucleus via the local rearrangement between Sm and Al atoms, while others are incorporated into the nucleus. The dominant role of the interfacial attachment mechanism might not hold true for nucleation at high Sm concentration or in the crystal growth regime (past the critical nucleus radius). In those cases it is possible that a Sm-rich shell is formed around Al nucleus/crystal interface and thereby the diffusion of Sm away from the nucleus/crystal interface might be expected to play the dominant role, as has been suggested in Refs.~\cite{wilde1999glass, imhoff2012kinetic}. Further studies are needed to confirm the role of solute diffusion in the nucleation kinetics for high solute concentrations and for longer time.

Although it has been shown that Al nucleation attachment occurs highly heterogeneously via collective movements of a group of atoms, such a collective attachment mechanism violates CNT which assumes the nucleus growth is continuous via single atom attachment. However, the application of CNT to extract the nucleation kinetic barrier $Q$ in Eq.~\ref{eq_kinetic_controlled} can still be justified, since $Q$ would represent the \textit{average} kinetic barrier per atom in the collective attachment.

\section{Conclusion}

The isothermal nucleation kinetics in Al-Sm MG was studied with $x_\mathrm{Sm}$ in the range of 0.0 - 3.0~at.\%. The following general conclusions can be drawn from the results of our MD simulations:
\begin{enumerate}[(1)]
	\item Sm solutes significantly suppress the primary crystallization transformation in Al-Sm MG and thereby they enhance its glass formability. In particular, the nucleation delay time increases exponentially with $x_\mathrm{Sm}$ at a constant $T/T_\mathrm{g}$ ratio, whereas the critical cooling rate $R_\mathrm{c}$ decays exponentially with $x_\mathrm{Sm}$. In addition, a reasonable range of $R_\mathrm{c}$ ($2.24 \times 10^4$ - $1.59 \times 10^6$~K/s) has been predicted for Al$_{92}$Sm$_8$, which is a well-known marginal glass former.
	\item Sm solute retards Al nucleation by increasing the nucleation kinetic barrier (and therefore by reducing the nucleus attachment rate). In the composition range considered in this study, the nucleation kinetics is controlled by interfacial attachments, rather than by Sm diffusion. 
	\item Similarly to Al diffusion in MGs, attachment of Al to a crystalline nucleus takes place heterogeneously in both space and time via the collective movement of a group of atoms. The common atomic-level processes between Al attachment and Al diffusion is consistent with the two processes having comparable activation energy barriers.
	
\end{enumerate}

\section*{Acknowledgments}
The authors are grateful to Dr. Ye Shen for helpful discussions of the critical cooling rates, and the Center for High Throughput Computing at University of Wisconsin, Madison for the computing resources. This work was supported by the NSF-DMREF under Grants DMR-1332851 and DMR-1728933.

\bibliographystyle{ActaMatnew-2}
\bibliography{AlSm_Nanocrystallization}

\end{document}


\begin{flushleft}
	\center{\textbf{\Large Supplemental Materials for ``Nucleation Kinetics in Al-Sm Metallic Glasses"}}
	\center{L. Zhao, G.B. Bokas, J.H. Perepezko, I. Szlufarska*}
	\vspace{-10pt}
	\center{\textit{Department of Materials Science and Engineering, University of Wisconsin - Madison, United States}}
\end{flushleft}

\section{Temperature history and nucleation simulation snapshots of Al-Sm samples}

\begin{figure}[htb]
	\centering
	\vspace{-10pt}
	\includegraphics[scale = 0.45]{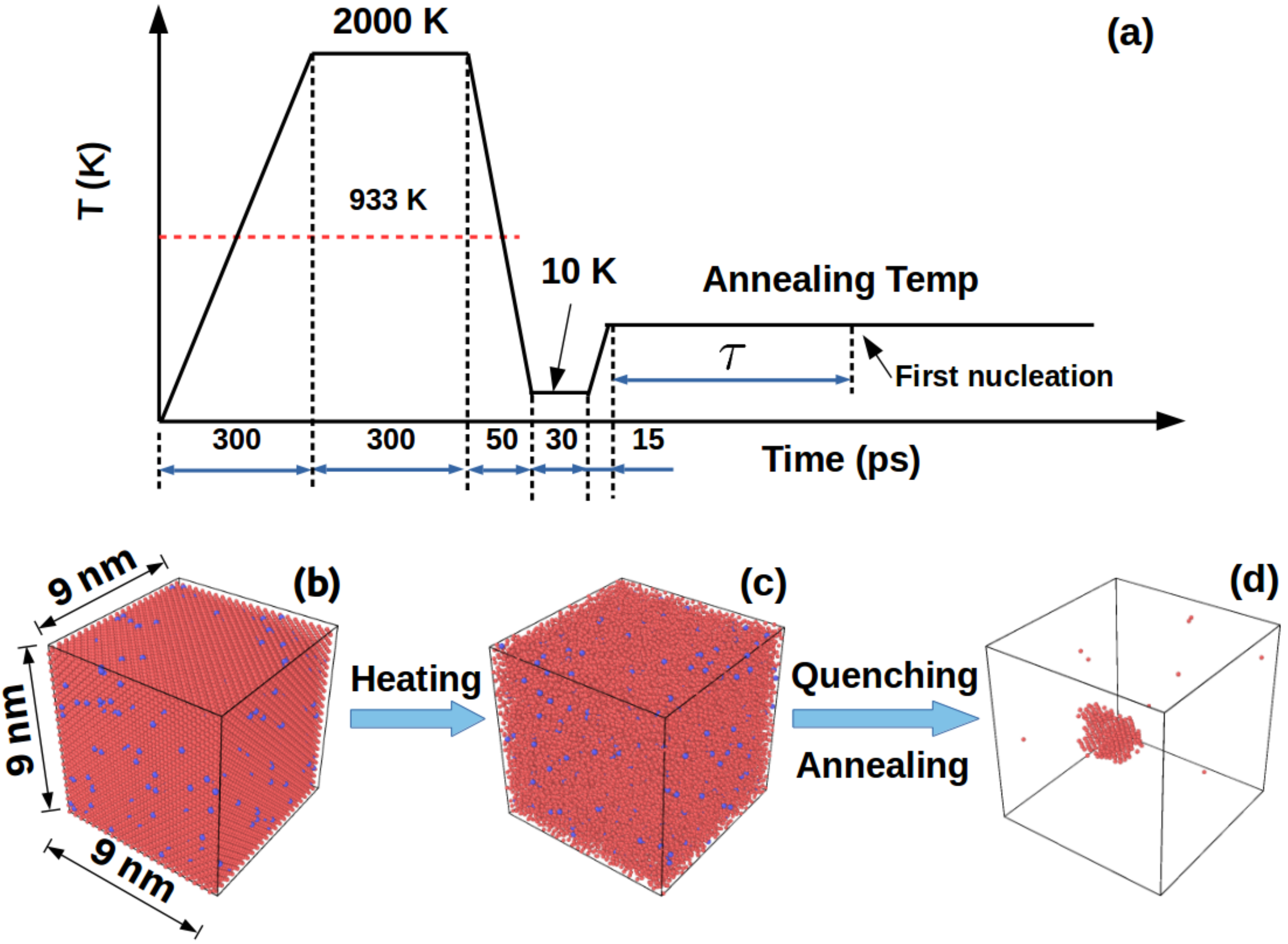}
	\vspace{-2pt}
	\caption{(Color online) (a) The temperature history of Al-Sm samples. Delay time $\tau$ of the first nucleation is recorded in each simulation. (b)-(d) Snapshots of nucleation simulation process. The red and blue atoms represent Al and Sm, respectively. Only Al FCC atoms are shown in (d).}
	\label{fig_simulation_cell}
\end{figure}

The temperature history and the snapshots from simulations of heating, quenching and annealing are shown in Fig.~\ref{fig_simulation_cell}. In the molecular dynamics (MD) simulations, it is observed that the face-centered cubic (FCC) Al crystal nucleates directly from the glassy state without the formation of any intermediate structures. This observation is consistent with those from the atomistic simulations of solidification of pure Al~\cite{desgranges2007molecular, hou2016cooling}, and is distinct from the reported nucleation mechanism in Lennard-Jones fluids and some metallic systems where a metastable BCC structure acts as the precursor for nucleation of the stable FCC phase~\cite{wendt1978empirical,ten1996numerical,zhou2017molecular}.

\section{Delay time determination}

According to classical nucleation theory (CNT)~\cite{walton1969nucleation, markov2003crystal}, the nucleation delay time $\tau$ is the time required by the new phase cluster to grow and overcome the nucleation energy barrier. In experiments, $\tau$ is usually determined by fitting the measured number of nuclei as a function of annealing time. This approach gives the average delay time over a number of nuclei. Here, however, $\tau$ is only recorded for the first nucleation event, due to the temporal and spatial scale limitations of MD simulations. 

There are two different approaches to determine $\tau$ of the first nucleus in the simulated samples. The first approach is to treat $\tau$ as the time the first nucleus takes to reach the critical nucleus radius $r_\mathrm{c}$. This approach is accurate once the exact value of $r_\mathrm{c}$ is known and therefore it requires that $r_\mathrm{c}$ is determined first. To determine a single $r_\mathrm{c}$ at each temperature and composition from MD simulation, one typically needs to embed clusters of different sizes into an amorphous matrix to observe whether they would grow or shrink at that annealing temperature. For each size of the embedded cluster, one should perform multiple independent simulations to obtain statistically meaningful results. Consequently, it becomes computationally expensive to determine all the $r_\mathrm{c}$ for multiple temperatures and compositions. In addition, the embedded cluster is spherical, while the true nucleus usually has preferred surface planes. Since the effect of the cluster morphology on $r_\mathrm{c}$ is not considered in this approach, it may lead to some inaccuracies in the estimates of $r_\mathrm{c}$ as well as perhaps $\tau$. The second approach to determine $\tau$ is simply based on the direct observation of the occurrence of the first nucleus. In the simulations the majority of time $\tau$ is spent on the formation of pre-nuclei and their sluggish growth. However, once the nucleus forms, it is found to grow at a relatively high rate. This transition of growth rate is used as a criterion to estimate the nucleation delay time $\tau$.

\begin{figure}[h]
	\centering
	\includegraphics[scale = 0.6]{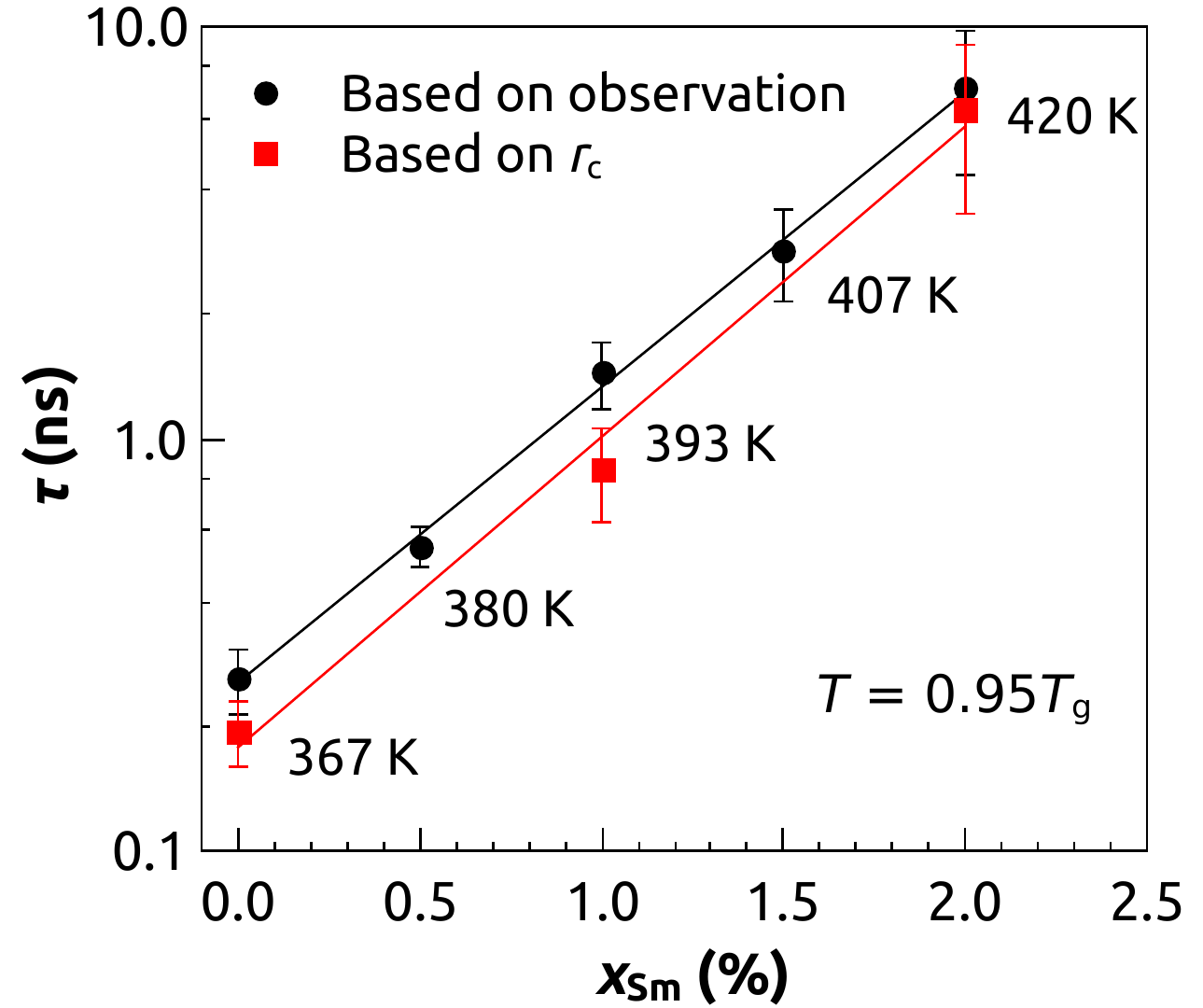}
	\vspace{-2pt}
	\caption{(Color online) Comparison of nucleation delay time $\tau$ determined by observation and critical nucleus radius $r_\mathrm{c}$. The annealing temperatures $T$ are set to $0.95\,T_\mathrm{g}$ and are shown along the fitted lines, where $T_\mathrm{g}$ is the respective annealing temperature for any composition $x_\mathrm{Sm}$. The error bar represents the standard deviation of $\tau$ from 10 independent simulations.}
	\label{fig_delay_time_determination}
\end{figure}

In order to compare the two aforementioned approaches, the delay time $\tau$ are determined from both methods for different Sm concentrations $x_\mathrm{Sm}$ at a fixed fraction (0.95) of glass transition temperatures $T_\mathrm{g}$. The results are shown in Fig.~\ref{fig_delay_time_determination}, where one can see that the two approaches produce comparable delay times. Therefore, for the sake of computational efficiency all the delay times in the main text are determined using the second approach, i.e., through the direct observation of occurrence of the first nucleus.

\section{Steady state nucleation rate}

\begin{figure}[h]
	\centering
	\includegraphics[scale = 0.6]{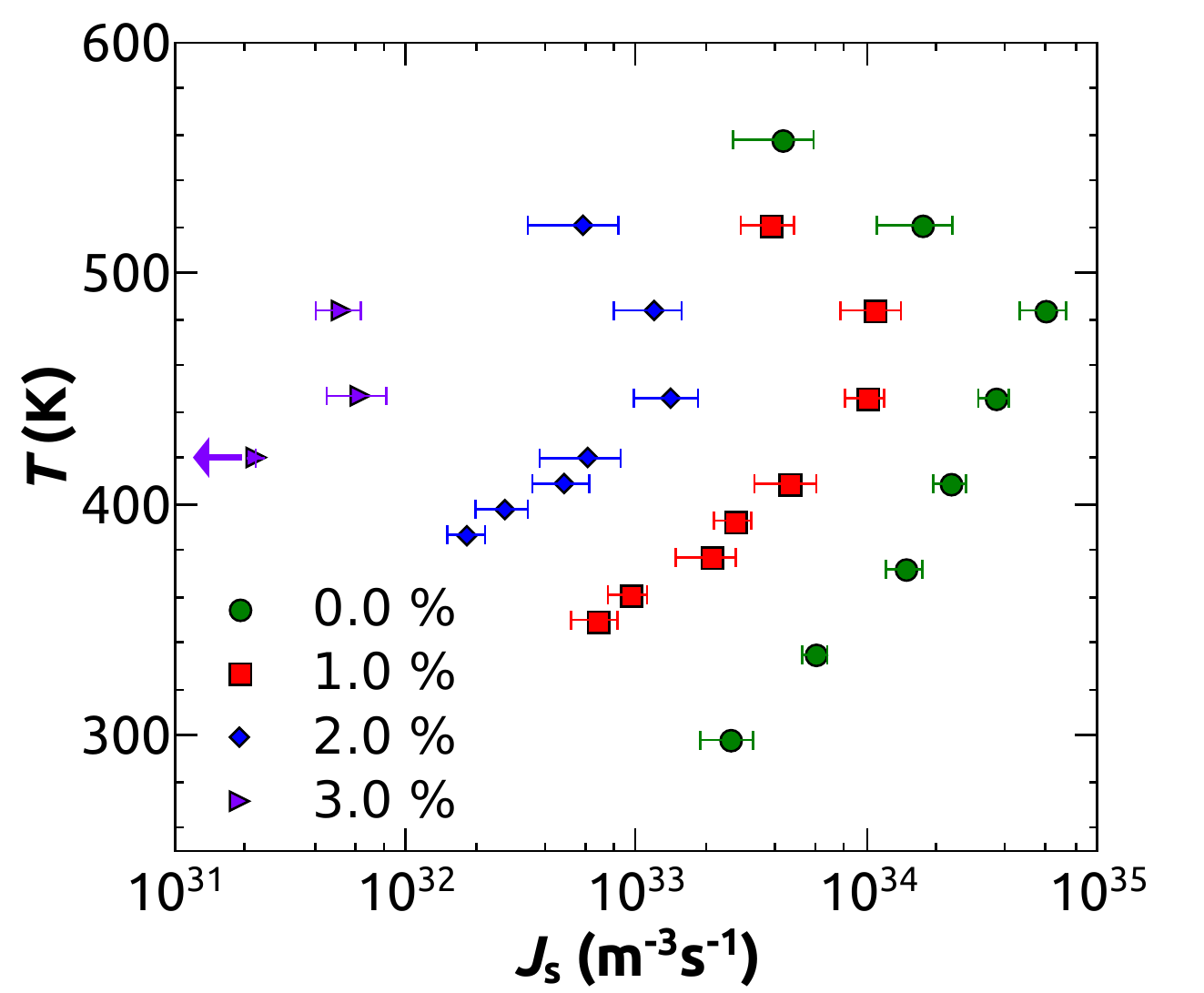}
	\vspace{-2pt}
	\caption{(Color online) The steady state nucleation rate $J_\mathrm{s}$ at different temperatures for different Sm concentrations.}
	\label{fig_nucleation_rate}
\end{figure}

To calculate the steady state nucleation rate $J_\mathrm{s}$, it is assumed that the instantaneous nucleation rate at $t = \tau$ is approximately $1/(V\tau)$, where $V$ is the system volume. Unfortunately, this assumption cannot be verified in this work because it is difficult to obtain the statistics of delay times of the initial a few (at least two) nuclei due to the time scale limitations of MD simulations. Plugging in the nucleation rate $1/(V\tau)$ at $t = \tau$ into the instantaneous nucleation rate equation $J(t) = J_\mathrm{s} \exp(-\tau/t)$, one can obtain the estimated steady state nucleation rate $J_\mathrm{s} \approx e/(V\tau)$. Fig.~\ref{fig_nucleation_rate} shows the estimated $J_\mathrm{s}$ at different temperatures, which (by construction) shows the opposite trend to the delay time in Fig.~\ref{fig_ttt_curves} in the main text.

\section{Calculation of diffusion energy barriers}

The diffusion coefficient $D$ is calculated as $D = \langle\left| \mathbf{r}(t) - \mathbf{r}(0)\right|^2 \rangle/6t$, where $\mathbf{r}$ is the position of a diffusing atom and $t$ is the time interval measured during annealing before the occurrence of the first nucleation event. The brackets $\langle \rangle$ represent an ensemble average, which is calculated as an average over all atoms of a given species in the system. Considering that MGs often exhibit temperature dependent diffusion energy barriers~\cite{fielitz1999diffusion, knorr1999self}, the diffusion coefficients are calculated for $\geq$~1.0~at.\% Sm in a wide temperature range below $T_\mathrm{nose}$. One exception is the pure Al system where the nucleation delay time for temperatures $T > 360$~K is too short for the steady-state (and therefore for a statistically meaningful) mean squared displacements to be obtained. Therefore, for pure Al system the diffusion coefficient is calculated only in a lower temperature range.

\begin{figure}[h]
	\vspace{20pt}
	\centering
	\captionsetup[subfigure]{labelformat=empty}
	\begin{subfigure}{0.32\textwidth}
		\includegraphics[scale = 0.40]{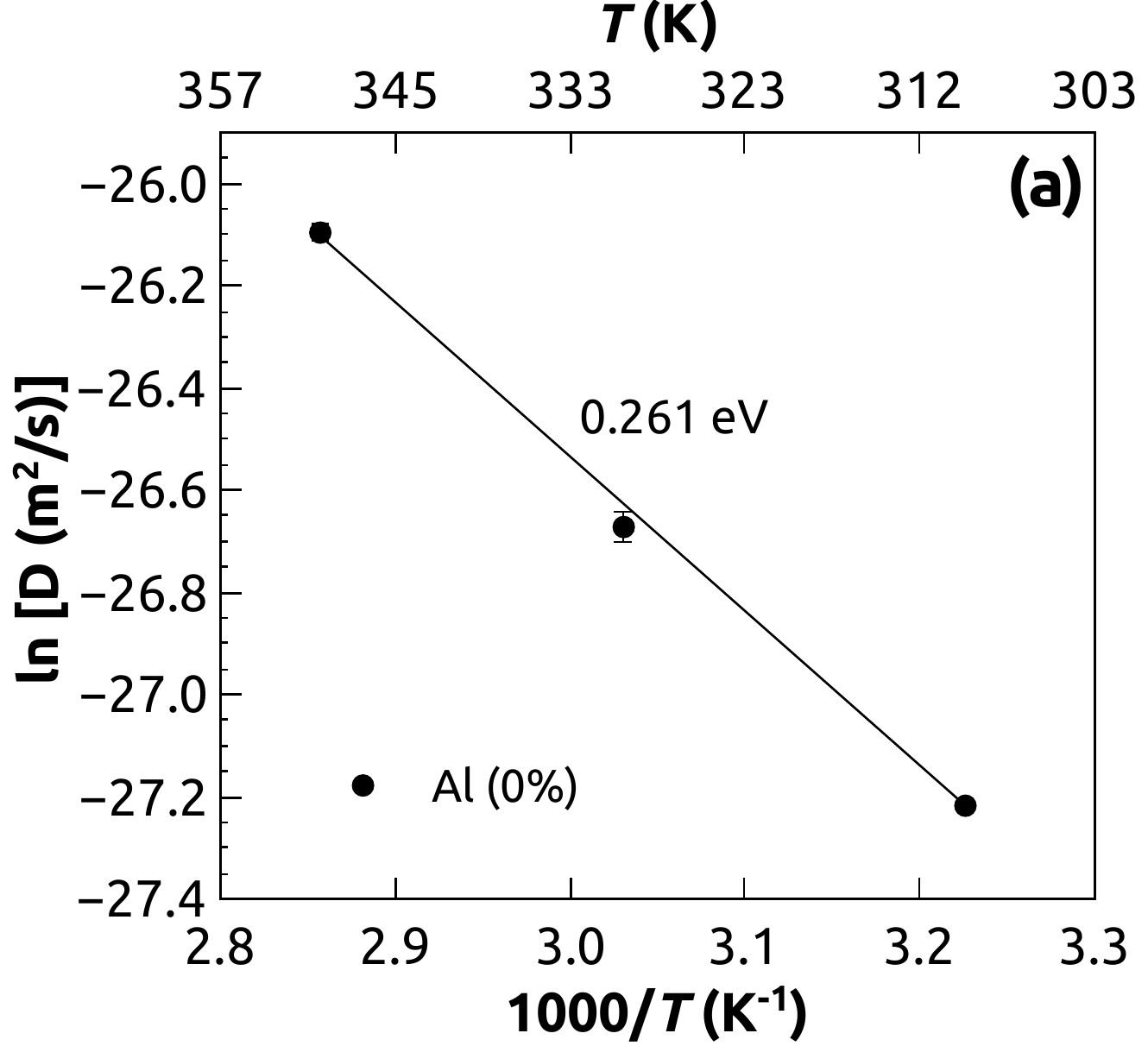}
		\caption{\centering}
		\vspace{-5pt}
		\label{fig_0pt_diffusion}
	\end{subfigure}
	\thinspace\thinspace 
	\begin{subfigure}{0.32\textwidth}
		\includegraphics[scale = 0.40]{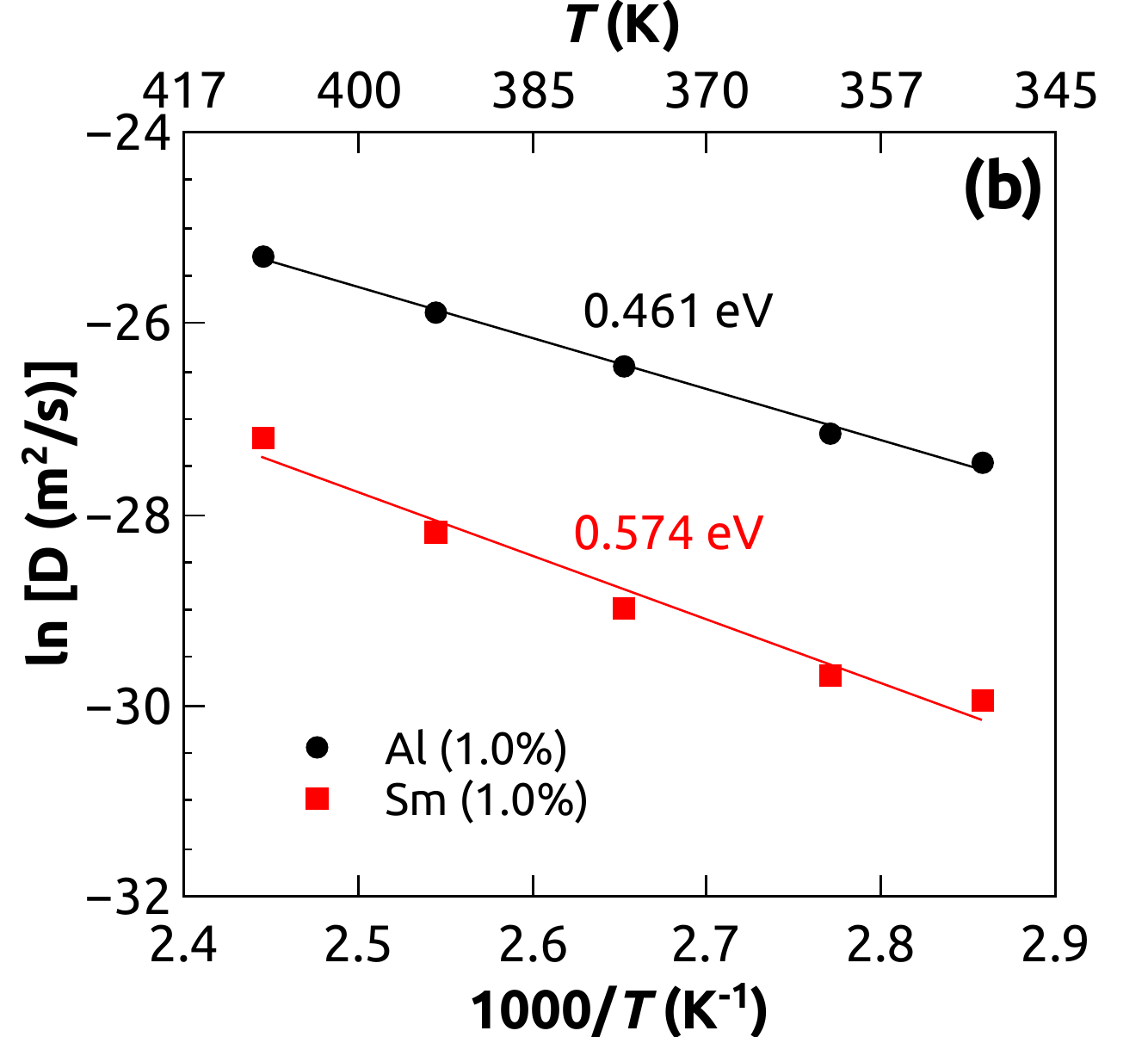}
		\caption{\centering}
		\vspace{-5pt}
		\label{fig_1pt_diffusion}
	\end{subfigure}
	\thinspace
	\begin{subfigure}{0.32\textwidth}
		\includegraphics[scale = 0.40]{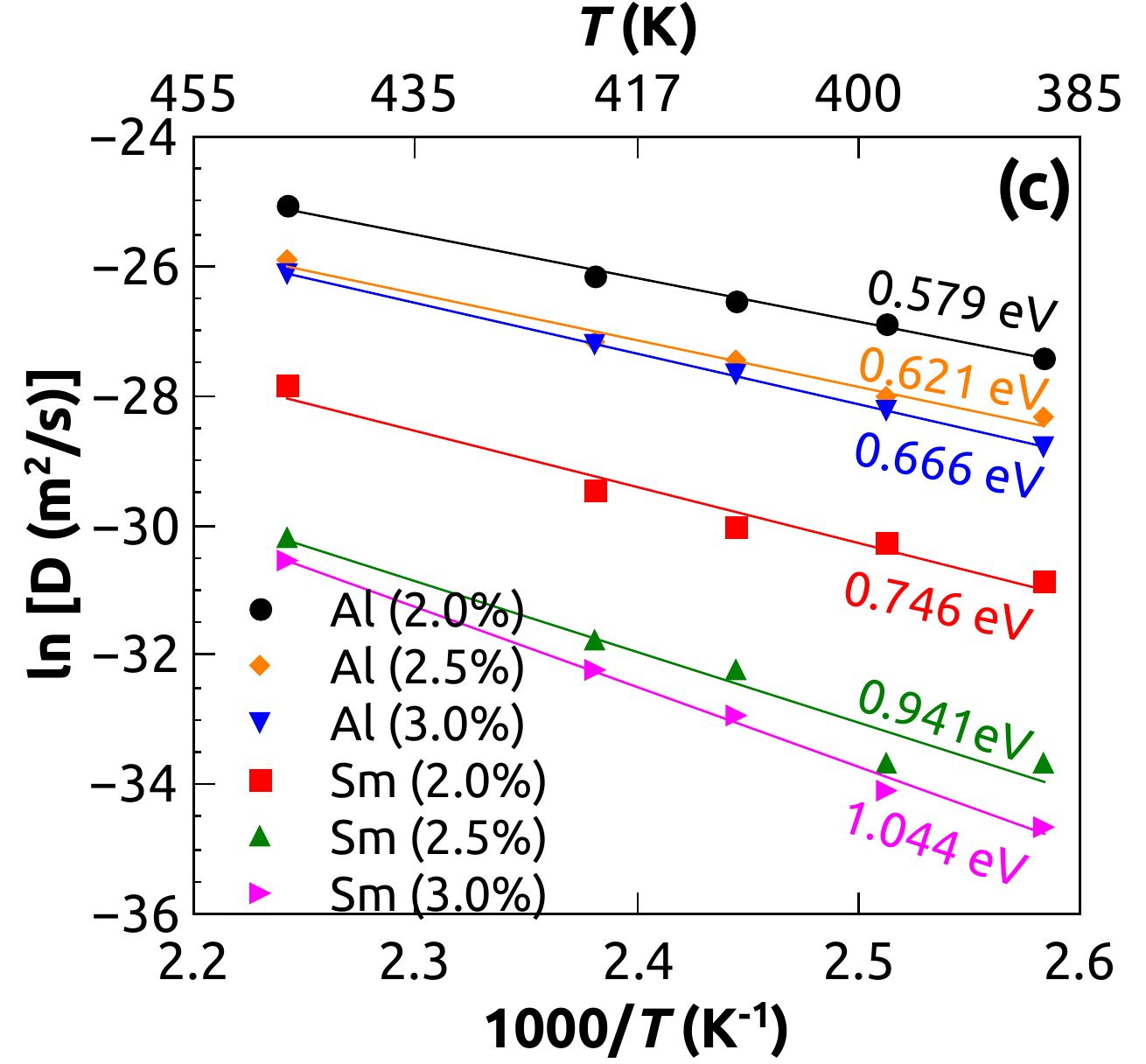} 
		\caption{\centering}
		\vspace{-5pt}
		\label{fig_2pt_diffusion}
	\end{subfigure}
	\vspace{-20pt}
	\caption{(Color online) Diffusion coefficient $D$ as a function of inverse temperature and fits to the data (lines) based on the Arrhenius relations for (a) $x_\mathrm{Sm}$ = 0.0~at.\%, (b) $x_\mathrm{Sm}$ = 1.0~at.\% and (c) $x_\mathrm{Sm}$ = 2.0, 2.5 and 3.0~at.\%, respectively.}
	\label{fig_diffusion_activation}
\end{figure}

The Arrhenius equation, $D = D_0 \exp(-E/k_\mathrm{B}T)$, is assumed to be followed, where $D_0$ and $E$ are diffusion prefactor and energy barrier, respectively. $D$ as a function of the inverse temperature and the Arrhenius fitting lines are shown in Fig.~\ref{fig_diffusion_activation}. The values of $D_\mathrm{0}$ and $E$ are obtained from the Arrhenius fitting and are listed in Table~\ref{tab_diffusion}. From Fig.~\ref{fig_diffusion_activation} one can see that the diffusion coefficients of both Al and Sm decrease and their energy barriers increase with increasing Sm concentration. This analysis suggests that Sm addition leads to a sluggish atomic mobility and to an improved structural stability of Al-Sm MG.

\begin{table}[h]
	\centering
	\vspace{-4pt}
	\begin{tabular}{ccccc}
		\toprule
		\multirow{2}{*}{$x_\mathrm{Sm}$(at.\%)}	& \multicolumn{2}{c}{$D_\mathrm{0}$\,(m$^2/$s)}	& \multicolumn{2}{c}{$E$\,(eV)}	\\ \cmidrule{2-5}
				& Al					& Sm					& Al		& Sm 		\\ \midrule
		0.0		& $2.6\times10^{-8}$	& 						& 0.261		&			\\
		1.0		& $4.8\times10^{-6}$	& $1.5\times10^{-5}$	& 0.461		& 0.574		\\
		2.0		& $4.3\times10^{-5}$	& $1.8\times10^{-4}$	& 0.579		& 0.746		\\
		2.5		& $5.3\times10^{-5}$	& $3.1\times10^{-3}$	& 0.621		& 0.941		\\
		3.0		& $1.9\times10^{-4}$	& $5.5\times10^{-2}$	& 0.666		& 1.044		\\
		\bottomrule
	\end{tabular}
	\caption{Fitted values of diffusion prefactor $D_\mathrm{0}$ and energy barrier $E$.}
	\label{tab_diffusion}
\end{table}

\section{Diffusing atoms}

\begin{figure}[h]
	\centering
	\includegraphics[scale = 0.6]{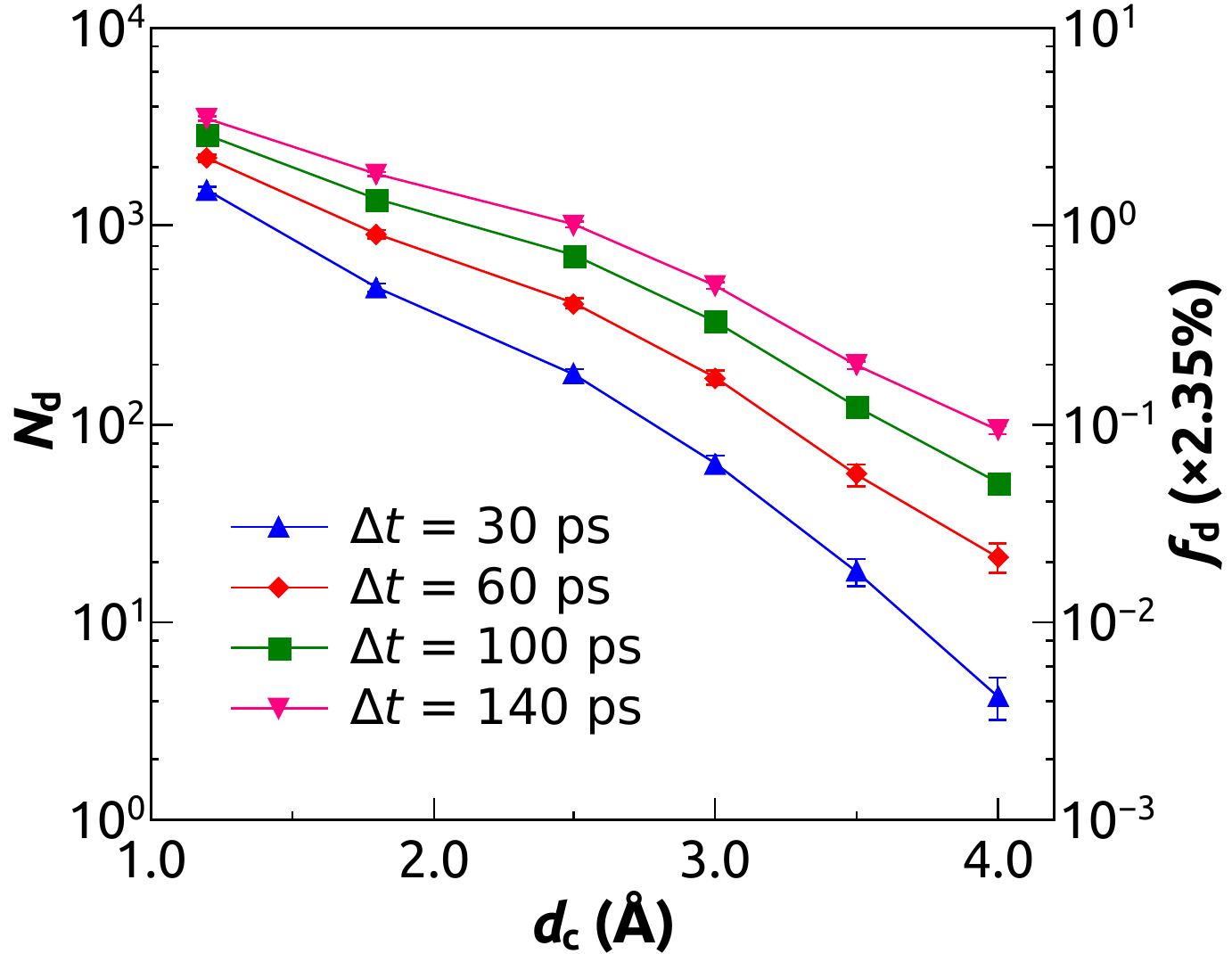}
	\vspace{-2pt}
	\caption{(Color online) Number $N_\mathrm{d}$ and fraction $f_\mathrm{d}$ of DAs as functions of the displacement cut-off $d_\mathrm{c}$ and the time window $\Delta t$ for a relaxed system with $x_\mathrm{Sm}$ = 2.0~at.\% and $T = 0.95\,T_\mathrm{g}$.}
	\label{fig_Num_DiffAl}
\end{figure}

The DAs are defined as atoms that undergo displacements larger than a cut-off $d_\mathrm{c}$ within a time window $\Delta t$. Fig.~\ref{fig_Num_DiffAl} shows the number $N_\mathrm{d}$ and fraction $f_\mathrm{d}$ of diffusing atoms (DA) for different displacement cut-offs $d_\mathrm{c}$ and time windows $\Delta t$. As expected, $N_\mathrm{d}$ and $f_\mathrm{d}$ decrease with increasing $d_\mathrm{c}$ and decreasing $\Delta t$. $f_\mathrm{d}$ is usually lower than a few percent. The small fraction of DAs indicates that majority of atoms only perform localized vibrations.

\vspace{50pt}
\bibliographystyle{ActaMatnew-2}
\bibliography{AlSm_Nanocrystallization}